\documentclass{aa}

\usepackage{graphicx}
\usepackage{amsmath,epsfig,rotating,amssymb}

\begin{document}

\def \rot{{\rm {\bf rot} }}
\def \grad{{\rm {\bf grad} }}
\def \div{{\rm div}}
\def \cha{\widehat}
\def \pr{{\it permanent}  regime }


\author{Audit E. \inst{1} and Hennebelle P.\inst{2}}

\institute{  Laboratoire AIM, CEA/DSM-CNRS, universit\'e Paris Diderot, IRFU/SAp, 
  \newline    91191 Gif-sur-Yvette Cedex
\and
         Laboratoire de radioastronomie millim{\'e}trique, UMR 8112 du CNRS, 
\newline {\'E}cole normale sup{\'e}rieure et Observatoire de Paris, 24 rue Lhomond,
\newline 75231 Paris cedex 05,
France}

\offprints{E. Audit, P. Hennebelle  \\
{\it e-mail:} edouard.audit@cea.fr, patrick.hennebelle@ens.fr}   

\title{On the Structure of the Turbulent Interstellar Clouds. }
\subtitle{Influence of the equation of state  on the Dynamics of 3D Compressible Flows}

\titlerunning{On the 3D Structure of Interstellar Clouds}

\abstract{It is well established that the atomic interstellar hydrogen
is  filling  the  galaxies  and  constitutes the  building  blocks  of
molecular clouds.}  {To understand  the formation and the evolution of
molecular clouds,  it is  necessary to investigate  the dynamics  of 
turbulent and  thermally bistable as well as barotropic flows.}  
{We  perform high resolution
3-dimensional  hydrodynamical simulations  of 2-phase,  isothermal and
polytropic  flows.}   {We compare  the  density  PDF  and Mach  number
density relation in the various simulations and conclude that 2-phase
flows  behave  rather  differently  than polytropic  flows.   We  also
extract the clumps and study their statistical properties such as mass
spectrum,  mass-size relation  and internal  velocity  dispersion.  In
each case,  it is found  that the behaviour  is well represented  by a
simple  powerlaw.   While  the  various exponents  inferred  are  very
similar  for   the  2-phase,  isothermal  and   polytropic  flows,  we
nevertheless find significant differences, such as for example the internal
velocity  dispersion  which  is  smaller  for  2-phase  flows.}   {The
structures statistics are very similar  to what has been inferred from
observations, in particular the  mass spectrum, the mass-size relation
and  the velocity  dispersion-size  relation are  all powerlaws  whose
indices well agree with the observed values.  Our results suggest that
in  spite  of  various   statistics  being  similar  for  2-phase  and
polytropic flows,  they nevertheless present  significant differences,
stressing the  necessity to consider  the proper thermal  structure of
the interstellar atomic hydrogen for computing its dynamics as well as
the  formation  of   molecular  clouds.}   

\maketitle

\begin{figure*}
\includegraphics[width=15cm,angle=0]{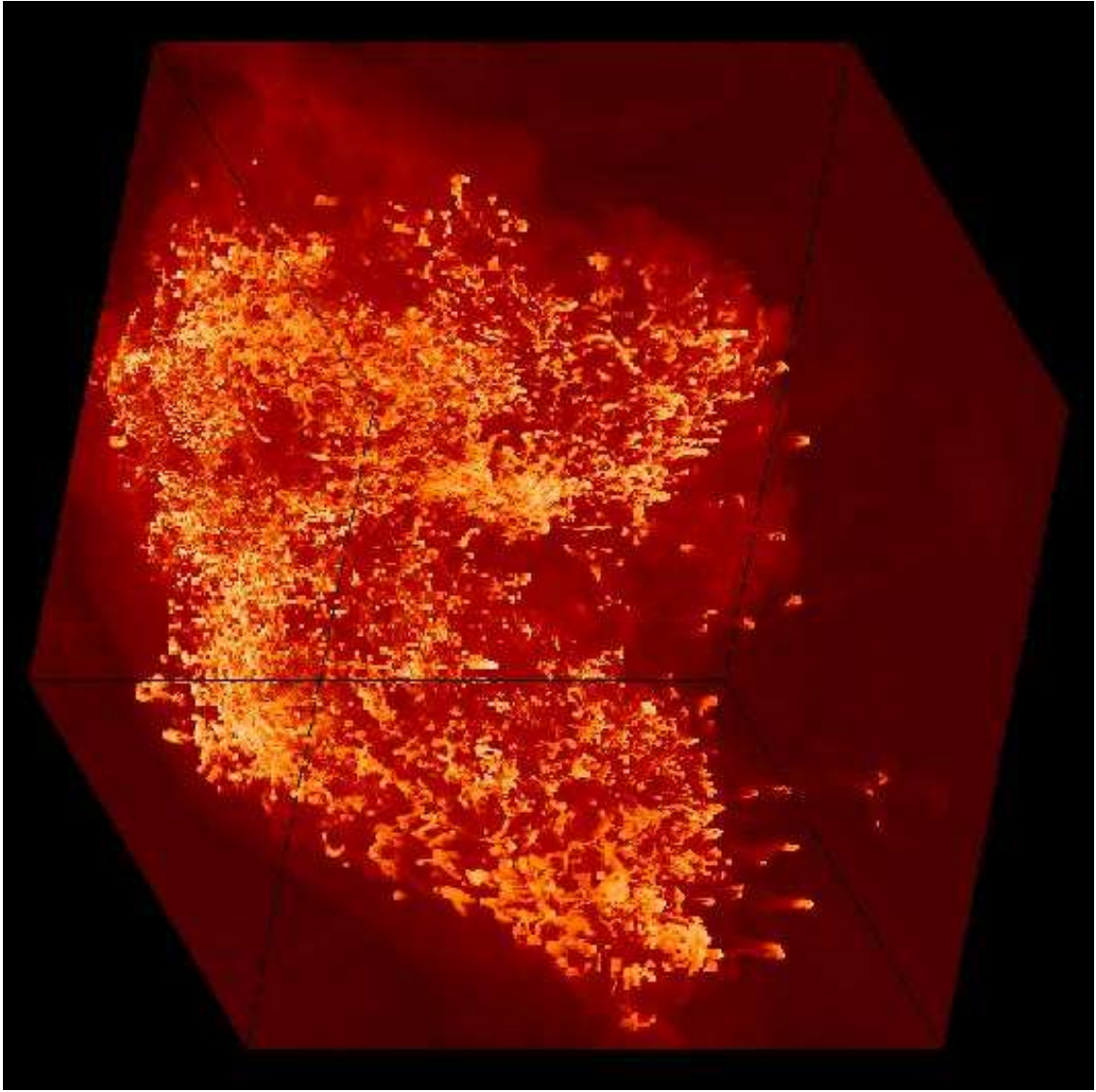}
\caption{Maximum density values  along the lines of sight  at one time
step  of the  $1200^3$  2-phase simulation.    The brighter  spots
correspond  to a density  exceeding 100  cm$-3$ embedded into a  pervasive
material at a density of a few cm$-3$. }
\label{3D1}
\end{figure*}

\begin{figure*}
\includegraphics[width=15cm,angle=0]{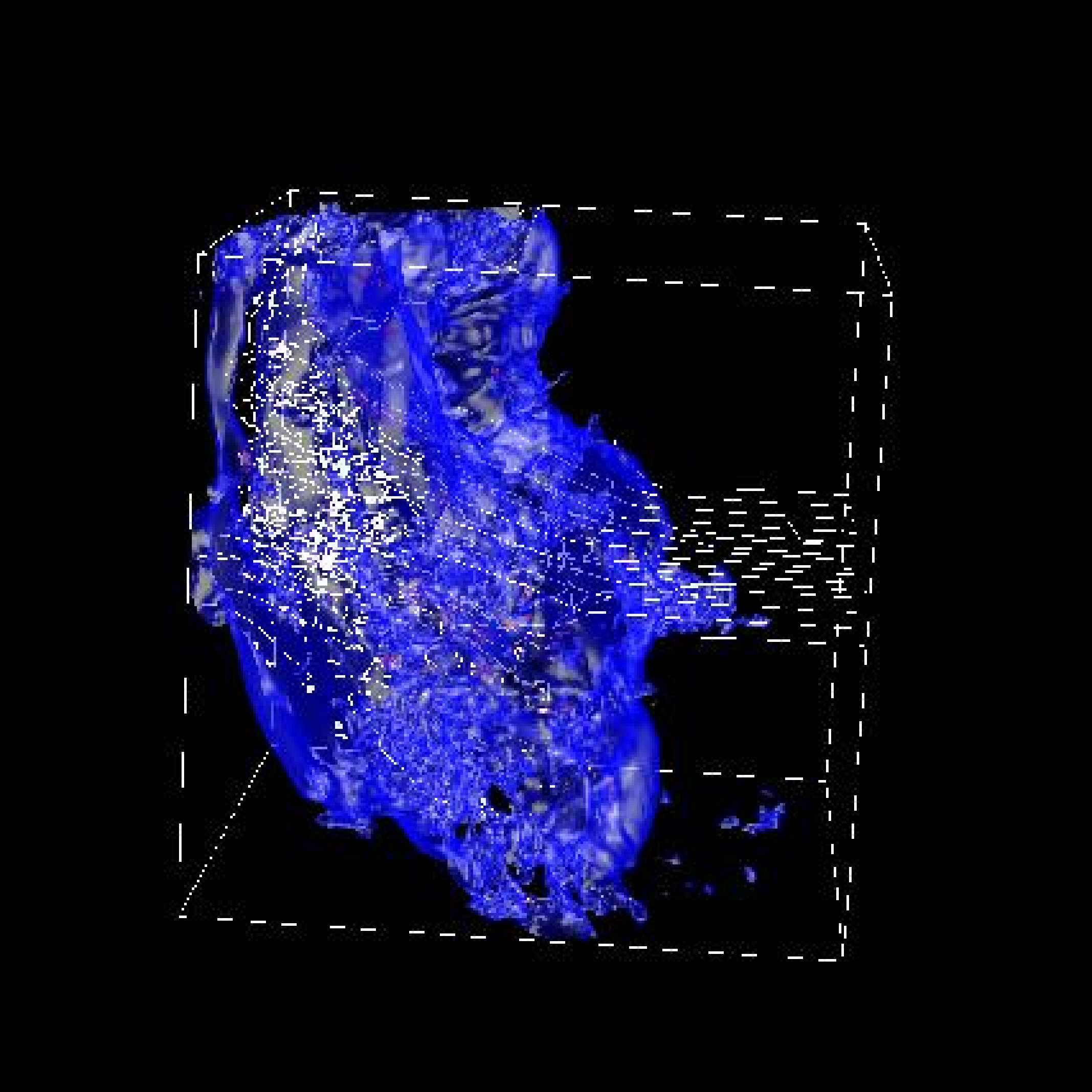}
\caption{Density isosurface and velocity streamlines at one time step of the 1200$^3$ 2-phase simulation. 
 The large isosurface corresponds to a density of 5 cm$^{-3}$ and the white clumps to a density of 500 cm$^{-3}$. 
One can see from the streamlines that the velocity fields is laminar in the WNM and becomes very turbulent
in and around the CNM structures.}
\label{bigchamps}
\end{figure*}

\section{Introduction}
Understanding the  interstellar turbulence  is of great  importance in
the  context of  molecular cloud  and  star formation.  As such,  many
theoretical  studies  and numerical  simulations  have been  performed
during the  last decades (see e.g.   the reviews by  MacLow \& Klessen
2004, Scalo  \& Elmegreen 2004 and  Elmegreen \& Scalo  2004). So far, most of
the  studies  have considered  isothermal flows.  Although this
constitutes  a reasonable  assumption  for the  densest  parts of  the
molecular  clouds,  it  is  not  an appropriated  assumption  for  the
description of  the interstellar atomic  hydrogen which is  2-phase in
nature (e.g., Dickey \& Lockman 1990, Field et al. 1969,  Wolfire et al.  1995) 
and therefore
for  the   formation  of  molecular  clouds.    Indeed,  recent  works
(Hennebelle \& Inutsuka 2006, V\'azquez-Semadeni et al. 2007, Heitsch et
al.  2008a,  Hennebelle  et  al.  2008,  Banerjee  et  al.  2009)  have
investigated  the possibility  that molecular  clouds  are multi-phase
objects. It seems therefore important to understand the dynamical properties of
a turbulent  2-phase medium as  the interstellar atomic  gas.  

Various
studies have been already performed  along this line. This includes 1D
calculations (Hennebelle  \& P\'erault 1999,  Koyama \& Inutsuka
2000,  S\'anchez-Salcedo  et al.  2002, Inoue et al. 2006),  
2D  calculations (Koyama  \&
Inutsuka 2002, Audit  \& Hennebelle 2005, Heitsch et  al. 2005, 2006) and 3D
calculations  (Kritsuk  \&  Norman  2002, Gazol et al. 2005, V\'azquez-Semadeni et al. 2006). 
The influence of the magnetic field
on the dynamics of a thermally bistable flow has been investigated by
Hennebelle \& P\'erault (2000), Piontek \& Ostriker (2004, 2005), 
de Avillez \& Breitschwerdt (2005), Hennebelle \& Passot (2006), 
and more recently by Inoue \& Inutsuka (2008), 
Hennebelle et al. (2009), Heitsch et al. (2009), 
 Inoue et al. (2009) and Gazol et al. (2009).

In  a  previous study,  Hennebelle  \&  Audit  2007, (hereafter  HA07)  and
Hennebelle et al. (2007), we have investigated the dynamics of 2-phase
flows  by  the  means   of  bidimensional  high  resolution  numerical
simulations.   In particular, we  have studied  the properties  of the
dense  and  cold  clumps  formed  out  of  the  warm  gas  by  thermal
instability, showing that they present many similarities with observed
clumps.  In the present paper, we investigate the properties of clumps
formed  in more realistic  3D simulations.   To better  understand the
influence  of the  2-phase  physics and since the nature of the inter-clump
gas within molecular clouds is not well known (see \S 4.1 
for a discussion), we  also  perform isothermal  and
polytropic simulations having identical setups as the 2-phase ones but
for the thermal  properties. We then compare the  results obtained for
the various types of simulations.

The  plan of  the paper  is the  following, in  the second  section we
describe the numerical experiments we perform, in the third section we
present the  global properties of the  numerical simulations including
the density  PDF and the Mach number-density  distribution. The fourth
section  is  devoted to  the  clump  properties  including their  mass
spectrum, velocity  dispersion and  mass-size relation.  In  the fifth
section we summarize our results and conclude the paper.

\begin{figure*}
\includegraphics[width=8cm,angle=0]{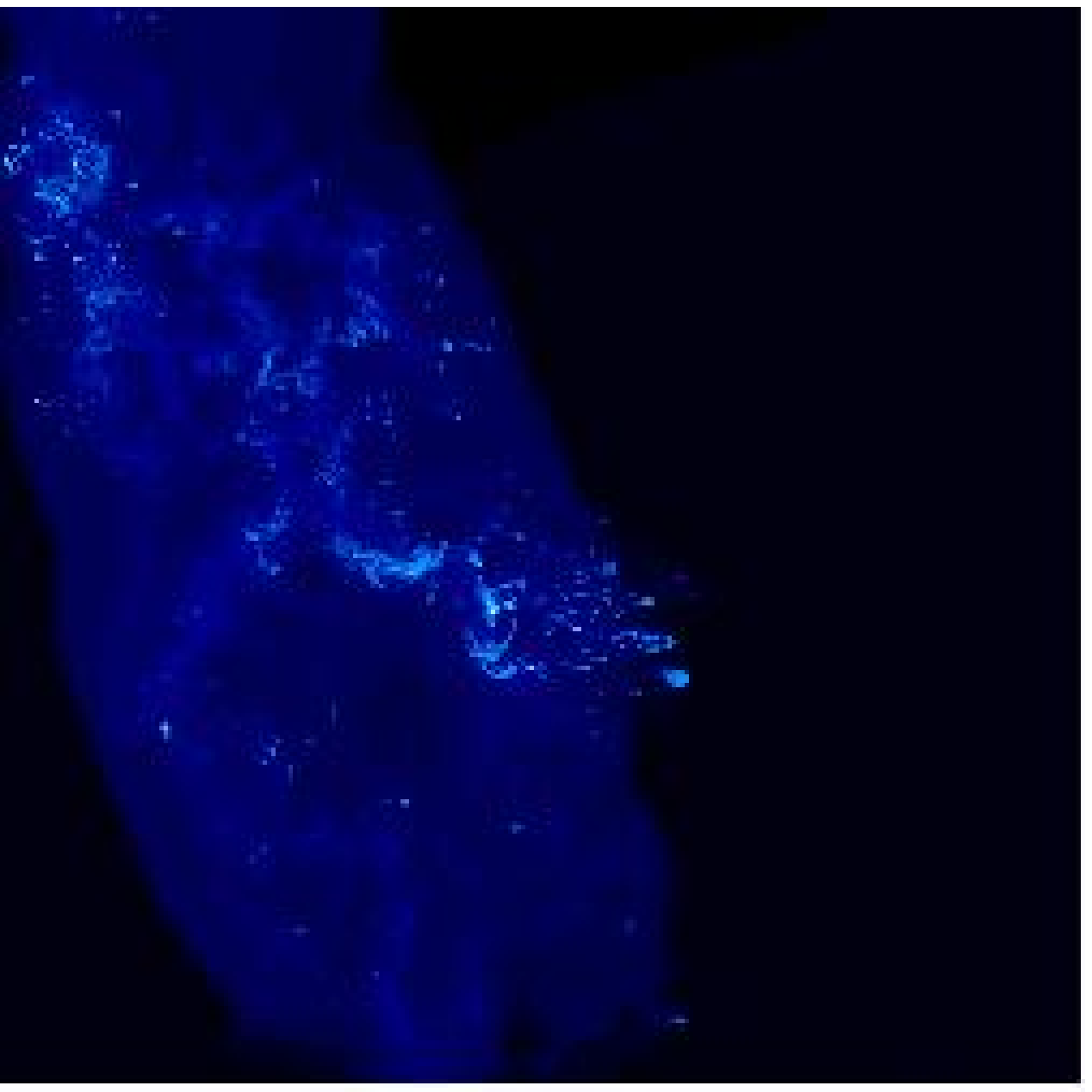}
\includegraphics[width=8cm,angle=0]{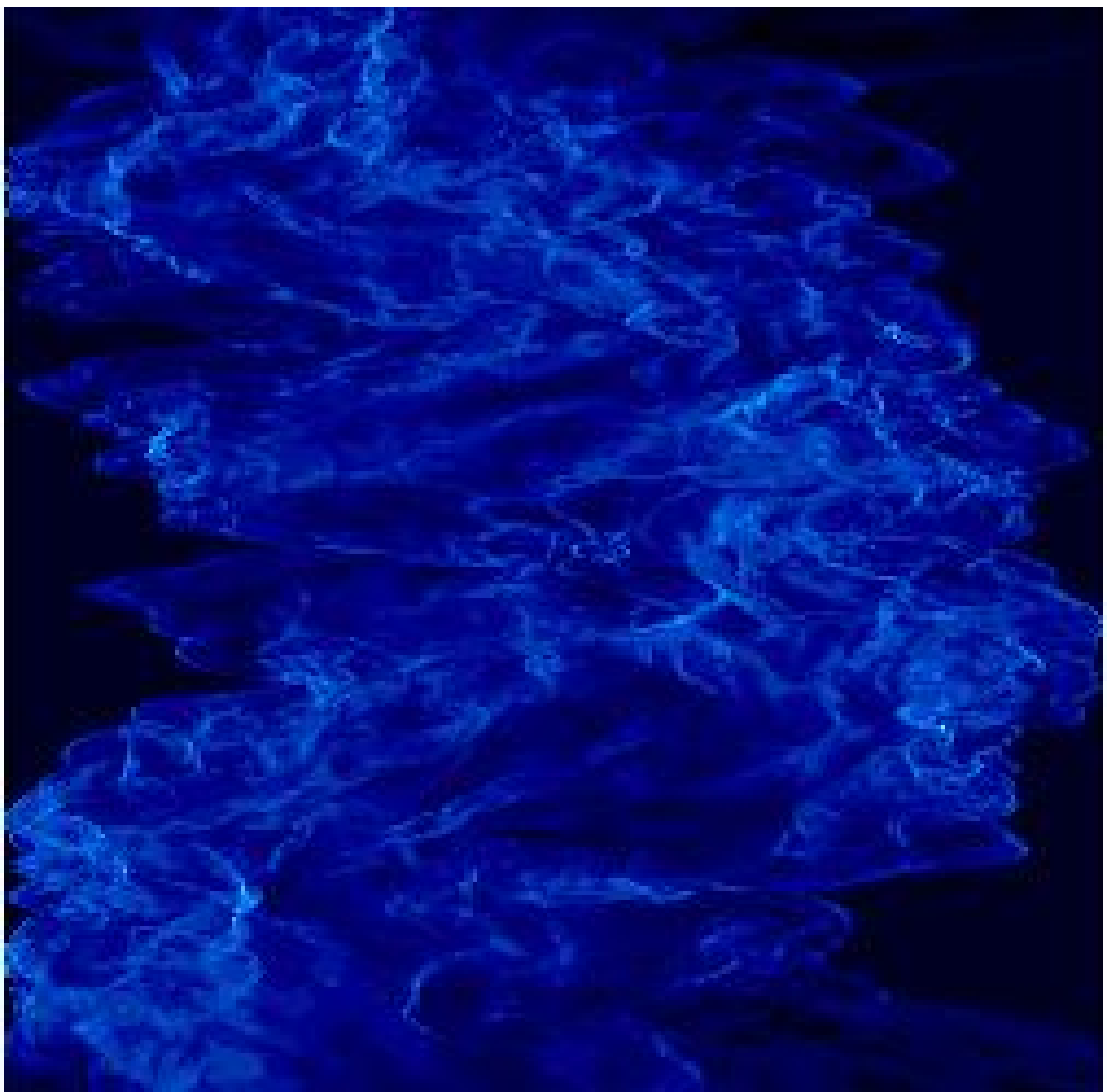}
\caption{Density   cut  through   the  simulations.   The   left  plot
  corresponds to the  2-phase run and the right  one to the isothermal
  run.  Bright  regions correspond to  a density between  20 cm$^{-3}$
  and 100  cm$^{-3}$ while  the pervasive grey  area corresponds  to a
  density of a few cm$^{-3}$.}
\label{coupes}
\end{figure*}

\section{Initial conditions and method}
\label{condini}

The equations are identical to  those used in our previous studies and
can be found in Audit \& Hennebelle (2005), (hereafter AH05). However,
one difference  with the  study performed  by HA07 is  that we  do not
include thermal conduction in the 3D runs. As emphasized in HA07, it does
not  have a major  effect except  on the  very small  scale structures
which  due  to insufficient  numerical  resolution,  are  not any  way
described in the present simulations.

\begin{figure}
\includegraphics[width=9cm,angle=0]{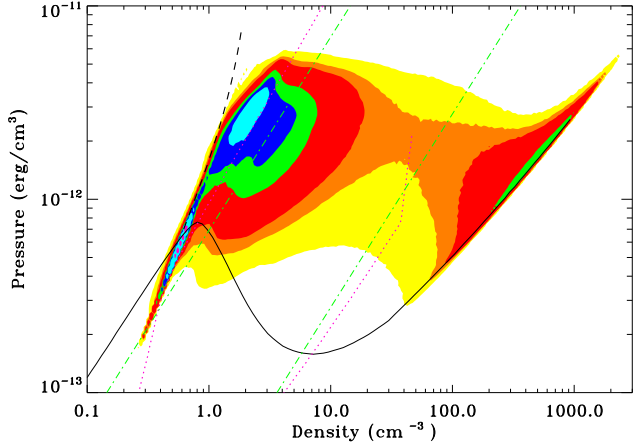}
\caption{Distribution of mass in the density-pressure plan.
The solid line corresponds to the thermal equilibrium curve, 
the dashed curve corresponds to the Hugoniot curve of shocked gas
and the dashed-dotted lines are the isothermal curves at $T=5000$ K
and $T=200$ K and the region between the dotted curves is the region where the gas is thermally unstable.   }
\label{press-dens}
\end{figure}



We use the HERACLES code to  perform the simulation.  This is a second
order Godunov type hydrodynamical code.  The size of the computational
domain is $15$  pc and the resolution ranges  from $600^3$ to $1200^3$
cells leading to a spatial resolution of $2.5 \times 10^{-2}$ to $1.25
\times  10^{-2}$  pc.   Let  us  remind  that  in  HA07,  the  highest
resolution  run  had  $10\;000^2$  cells corresponding  to  a  spatial
resolution  of  $2  \times  10^{-3}$  pc.   As  stated  in  HA07,  the
resolution  has  a  strong  influence  on  the  results  and  must  be
sufficient to  cover a large  dynamics of spatial  scales.  Typically,
2500$^2$ to 5000$^2$ cells were  needed to get some reliable numerical
convergence even if no strict numerical convergence could be obtained.
Therefore, the present runs are  likely to be affected by insufficient
numerical  resolution.   However,  we   stress  that  it  is  nowadays
difficult to  do much bigger runs in  3D than the one  we perform.  We
believe, nevertheless, that investigating the 3D effects is worthwhile
even with a relatively low resolution. One should however keep in mind
this restriction when looking at the results.

Since our primary goal is to  investigate the exact influence of  
equations of state on the dynamics of the flow,
 we  have    done  simulations  of 2-phase flows (using the cooling function
described in AH05) and using  an
isothermal equation  of state  with a temperature  of 100  K. Finally,
since it  has been  found that the  effective polytropic index  of the
cold  gas  is  about 0.7,  we  have  performed  a simulation  using  a
polytropic  equation of  state with  a polytropic  index of  $\gamma =
0.7$.  For this run the temperature is given by
\begin{eqnarray}
T = 100 \, {\rm K} \, ({ \rho \over 100 \, {\rm cm}^{-3} })^{-0.3}.
\end{eqnarray}

 For all the simulations, the boundary conditions
consist in an  imposed converging flow (see AH05, Folini \& Walder  2006)
at the left  and right faces of
amplitude   $V_{in} \simeq$15 km s$^{-1}$ 
(which corresponds to $\simeq 1.5 \times  C_{s,wnm}$ 
where $C_{s,wnm}$   is the sound  speed of  the  warm neutral
medium) on top  of which fluctuations of various  amplitudes have been
superimposed.  We stress that since the mean velocity is the same for 
all the runs, this implies that while 2-phase simulations are 
mildly transsonic (this is of course no longer the case for the cold gas 
produced in these simulations) with Mach number ${\cal M} \simeq 1.5$, the 
isothermal and barotropic ones are very supersonic with Mach numbers of 
the order of 10. 
The initial  conditions are  also identical to  the one used  in AH05,
that is to say a uniform low density gas of density $n_{\rm in} \simeq 0.8$ cm$^{-3}$
which is also the density of the incoming flow for all the runs.
In the 2-phase case,  this gas is at thermal  equilibrium and  corresponds  to  WNM
(i.e. has a temperature of about 8000 K).

For the 3 types of simulations, the ram pressure of the incoming flow 
is thus  $ n_{\rm in} m_p V_{in}^2 =  n_{\rm in} m_p \times (1.5 \times C_{s,wnm})^2 
\simeq 2.15 \times k_b n_{\rm in} T_{WNM} \simeq 1.4 \times 10^4 k_b$ K cm$^{-3}$, where 
$T_{WNM}$ is the temperature of the WNM which is of the order of 
8000 K, $k_b$ is the Boltzmann constant and $m_p$ the mean molecular weigth. 
In the 2-phase case, the thermal pressure of the diffuse gas injected inside the box is about 
$(3/2) k_b n_{\rm in} T_{WNM} \simeq 1.2 \times 10^4 k_b  $ K cm$^{-3}$ 
 while it is $(3/2) k_b n_{\rm in} \times (100 K) \simeq 150 k_b$ 
K cm$^{-3}$  in the isothermal case. 
The latter is thus roughly 80 times smaller than the former. 

For the four  other faces, outflow conditions have been
setup, implying that the flow  can escape the computational box across
these  four faces.   The advantage  of this  setup, which  consists in
focusing  on a  peculiar large  scale events,  is clearly  the spatial
resolution  of  the  CNM  structures.  In  particular,  considering  a
turbulent forcing  would require a  larger box to produce  large scale
events at  the scale of  the WNM cooling length,  therefore increasing
the size of the computational  cells. We stress that this is different
from   polytropic  flows in  which there  is no  cooling length  to be
considered.

The
simulations are then runned  until a statistically stationary state is
reached. This constitutes a  difference with other related work (e.g.,
Koyama \& Inutsuka  2002, Heitsch et al.  2006)  which investigate the
development of  the thermal and dynamical instabilities.   In order to
reach more rapidly the statistically stationary state, we start with a
coarser numerical  resolution and we  then double it  once statistical
stationarity has been reached.

In order  to study the  influence of various physical  parameters, and
more specifically  of the cooling function, we  have performed several
runs assuming different thermo-dynamical properties for the gas. First
of  all,  we have  done  three  runs with  a  2-phase  flow using  the
heating/cooling function presented in  AH05. For these three runs, the
amplitude  of the velocity  fluctuations imposed  on the  boundary was
given by  $\epsilon=1,2$ and $4$  corresponding to modulations  of the
order of 10  \%, 20 \% and 40 \% respectively  (see AH05 for details).
The effect of varying the amplitude of the fluctuations imposed at the
boundaries is to increase the  turbulence in the box. To quantify this
effect, we have computed the  mean shear for each value of $\epsilon$.
The average shear  is computed in the following way:  in each cell, we
compute the  velocity stress tensor (i.e.,  $\partial_{x_i} V_j$).  We
then diagonalize the symmetric trace-free  part of this tensor and the
shear   is  defined   as  the   root-mean  square   of   the  obtained
eigenvalues.  We finally  compute the  average over  all cells  in the
simulation.   Table~1   gives  the  value   of  the  mean   shear  for
$\epsilon=$1, 2 and 4. The  isothermal  run was  done  using  $\epsilon=2$  for the  boundary
conditions.

\section{Global statistics}

\subsection{General morphology of the flow}

Figure~\ref{3D1}  shows the  maximum of  the density  field  along the
lines  of   sight  at  a  particular  time   step,  after  statistical
stationarity has been reached in the box for the 2-phase run in the case
 $\epsilon=1$.   The morphology  is rather  complex. The
cold  gas  seen  in white  and  yellow  is  very fragmented  and  very
structured.  The warm and diffuse gas  appears to be intrusive. This
is qualitatively  similar to the structure observed  in 2D simulations
(see AH05 and HA07) although some differences can  be seen.  For example,
the 2D simulations seems to be more filamentary. It is worth, however,
to recall at  this stage that the resolution is 5  to 10 times smaller
in this run than in the highest resolution run presented in HA07.

Figure~\ref{bigchamps}   diplays  density  isosurfaces   and  velocity
streamlines  at  the  same  time  step.  The    large  isosurface
corresponds to  a density of  5 cm$^{-3}$ and therefore  traces mainly
the WNM  compressed by the ram  pressure of the  converging flow.  The
  white clumps  corresponds to  a  density of  500 cm$^{-3}$  and
therefore  mainly  shows the  dense  CNM  structures  confined by  the
thermal and the ram pressure  of the surrounding WNM.  The streamlines
show that the velocity field is  nearly laminar in the WNM and becomes
very turbulent  inside the compressed  layer and therefore  around the
CNM structures.   This is  similar to what  is reported by  Heitsch et
al. (2006, 2008b) who point  out the influence of the Kelvin-Helmholtz
instability  as a plausible  source of  turbulence within  the forming
cloud.

Figure~\ref{coupes} displays  density maps of  a slice of  the 2-phase
run (left) and of the isothermal one (right). The visual aspect of the
density field in  these two simulations is quite  different.  First of
all,  while the dense  isothermal flow  tends to  be distributed  in a
large fraction of  the box, most of the dense gas  in the 2-phase flow
is  concentrated  in a  thin  layer  located at  the  onset  of the  2
converging flows.  This is partly due to the thermal pressure which is
higher  in  the  2-phase case  than  in  the  isothermal gas  and  can
efficiently  push  the  gas  outside  the  computing  box  (indeed  as
discussed later the  total mass is higher in  the isothermal case than
in  the 2-phase  case).  However,  we have  also  performed runs  with
periodic boundary  conditions along the y and  z-directions which show
the same trends  although slightly reduced. Thus, we  think that other
dynamical processes  such as  the development of  the non  linear thin
shell instability (Vishniac 1994) is also playing a role here. Indeed,
for this instability to develop, the slab must be locally displaced by
at least  a distance comparable  to its thickness.  As  a consequence,
the non  linear thin shell instability  develops only, or  at least is
much stronger, when  the incoming flows are supersonic.   While in the
2-phase case, the  flows are transsonic (since the  sound speed of the
WNM is  about 10  km s$^{-1}$, they  are highly supersonic  (${\cal M}
\simeq 10$) in the isothermal case.

Second of all, in the
2-phase  case,  one sees  spheroid  structures,  with some  filaments
surounded  by a  more diffuse  medium.   In the  isothermal case,  the
density structures are  largely imprinted by bow shocks  and clumps can
hardly be  seen. As we will  see further the density PDF and Mach number 
distribution are indeed very different in both cases. However, 
 these  large 
differences are not so obvious  anymore when looking at the statistics 
of dense structures.

\subsection{Pressure and density distributions}

Figures~\ref{press-dens}   show   a  bidimensional   pressure-density
histogram.  The solid line is the thermal  equilibrium curve, the
two dashed-dotted lines correspond to constant temperature ($T=5000$ and 200 K) and the
dashed line corresponds to the  Hugoniot curve. 
As  in   previous  studies (V\'azquez-Semadeni et al. 2003, AH05), we see that
if  most of  the gas  lies  near the  two branches  of equilibrium  ($T \simeq
8000$ K for the WNM and $T \simeq 50-100$ K for the CNM),  a
significant fraction is nevertheless in the thermally unstable domain.
As  pointed  out in  AH05,  there  is a  clear
correlation between the level of turbulence 
  and the fraction of this thermally
unstable gas indicating that turbulence is the main cause for the existence of this gas.
This can be seen in table~\ref{table1}, where the fraction of cold and
unstable gas is given for each 2-phase runs as well as the mean value
of the shear. It is
clear that  the shear inhibits  the formation of cold  structures. For
the two  runs with the lowest  value of $\epsilon$ ($\epsilon = 1$  and $2$), we
find, as  in AU05,  that the amount of  cold and unstable gas is
independant  of  the  turbulence   but  that  the  formation  of  cold
gas is  strongly inhibited by the shear.  In other words, the 
shear does not prevent the formation of thermally unstable gas but 
does prevent the formation of cold dense gas out of the thermally unstable gas.

For $\epsilon=4$, the
shear is so strong that cold structures can hardly form. For this reason, in 
the following we emphasize the $\epsilon=1$ case which contains a larger fraction 
of CNM providing better statistics on the dense gas than the other cases. In the 
appendix the results corresponding to the $\epsilon=2$ case are displayed.

\begin{figure}
\includegraphics[width=9cm,angle=0]{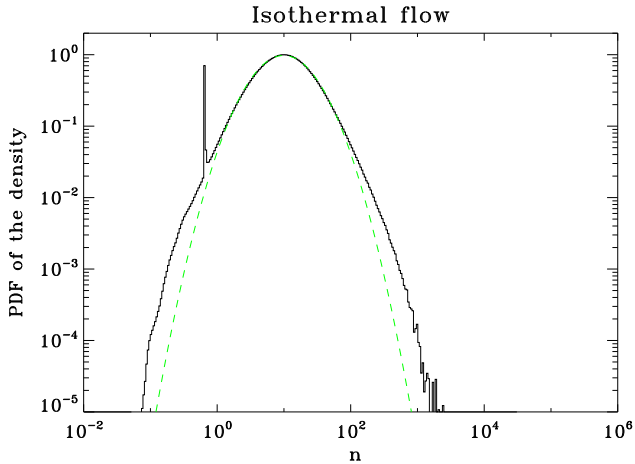}
\caption{Probability distribution function of the density for the isothermal run. The dashed line is a lognormal fit.}
\label{pdf2}
\end{figure}

\begin{figure}
\includegraphics[width=9cm,angle=0]{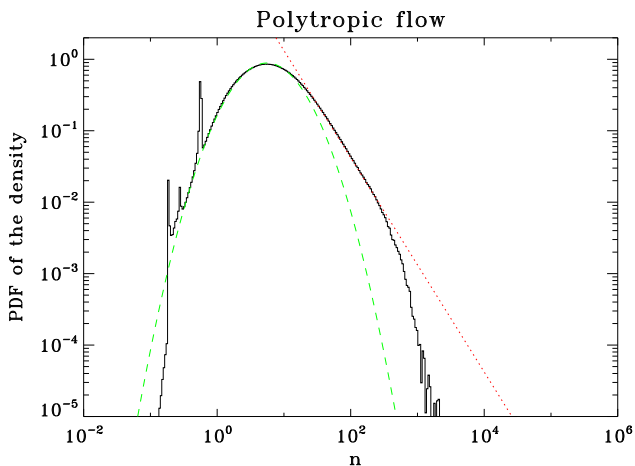}
\caption{Probability distribution function of the density for the polytropic run. The dashed line is a lognormal fit while
the dotted line is a powerlaw fit.}
\label{pdf3}
\end{figure}

\begin{figure}
\includegraphics[width=9cm,angle=0]{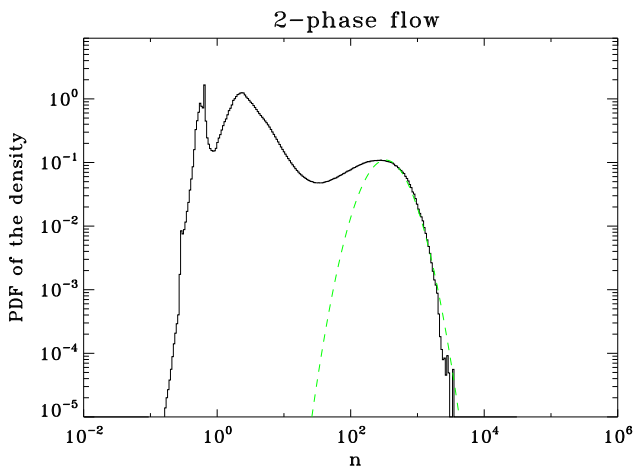}
\caption{Probability distribution function of the density for the 2-phase run.}
\label{pdf1}
\end{figure}

The  probality distribution  function  (PDF) of  the  density for  the
2-phase,  isothermal  and  polytropic  runs are  plotted  on  
Figs.~\ref{pdf2},  \ref{pdf3} and  \ref{pdf1}, respectively.   
The density PDF, $P(n)$, of isothermal simulations has been studied by various 
authors (e.g. Padaon et al. 1997, Passot \& V\'azquez-Semadeni 1998) 
and has been found 
to be lognormal. More precisely, the following formulae has been proposed:
\begin{eqnarray}
P(n) = \frac{a}{\sqrt{2\pi\sigma^2}} \exp(-\frac{( \ln(n)-\ln(\bar{n})-0.5\sigma^2)}{2\sigma^2})
\label{pdf}
\end{eqnarray}
with $\sigma^2 = \ln(1+b^2 {\cal M}^2 )$.
The fits shown in  the figures were obtained using  the parameters given
in table~2.  The  dotted line in Fig.~\ref{pdf3} corresponds
to $n^{-3/2}$.

As found by previous authors,  the isothermal PDF is well fitted
by this lognormal distribution, eventhough we get slightly larger wings.
This may be due to our forcing which is not applied in the Fourier space
and in the solenoidal modes as it is the case in most of the 
compressible simulations which have been performed. Indeed, 
Federrath et al. (2008) find that while forcing in the compressible modes rather than 
in the solenoidal ones, large non Gaussian wings develop. Since our 
forcing is exerted from the boundaries as a converging flow and therefore,
 at the scale of the box at least, mainly in the compressible modes, it seems
likely to us that such an effect is certainly playing a role in our simulation.
The value of the $b$ parameter, namely $b=0.33$ is also reminiscent of the values
quoted in the literature (e.g. Federrath et al. 2008).
 
In the $\gamma=0.7$  polytropic case, the low density  part of the PDF
is well fitted by a  lognormal distribution, while for higher density,
the PDF  is a  powerlaw whose exponent  is about -1.5.   Such powerlaws
have   been   numerically   found   and   explained   in   Passot   \&
V\'azquez-Semadeni  (1998) for 1D  simulations It  is typical  of very
compressible fluids ($\gamma < 1$)  and is a direct consequence of the
thermal   pressure   term.  Indeed,   for   $\gamma=0.5$,  Passot   \&
V\'azquez-Semadeni  (1998)  report an  exponent  of  about -1.2  which
appears to  be close to our  result.  Since the value  of the exponent
that we  inferred here  is in good  agreement with their  estimate, it
seems that the 3D effects are not altering their conclusion. Note that
the value  of the $b$ parameter  quoted in table~2  is only indicative
since as  seen from  Fig.~\ref{pdf3}, the lognormal  distribution does
not provide a good fit to the high density part of the PDF.

For the 2-phase run, we obviously  cannot get a lognormal PDF since by
definition,  the   PDF  is  bimodal.   However,   it  is  nevertheless
interesting to investigate  to what extend the cold  gas can be fitted
by a lognormal distribution. Table  2 gives the parameter values. Note
that  the Mach number  has been  estimated by  selecting computational
cells of density  larger than 20 cm$^{-3}$.  As can  be seen, the high
density  part  of the  cold  gas  density  distribution is  reasonably
reproduced  by a lognormal  fit.  This  is a  little surprising  to us
since the effective polytropic index of  the CNM is smaller than 1 and
indeed not  far from 0.7 as  can be seen  in Fig.~\ref{press-dens}. We
speculate  that  the  numerical   resolution  of  the  CNM  structures
available in  the present simulations,  is not sufficient  to describe
this behaviour.  We note however that the low density part of the CNM,
ranging  from $\simeq 10$  cm$^{-3}$ up  to the  peak at  $\simeq$ 300
cm$^{-3}$, is  not well described  by a lognormal  distribution.  This
clearly raises the  question of what density PDF  should be considered
in models of molecular clouds.

Finally, it is worth noting that the mean density, or equivalently the total 
mass, in the box is different for the 3 types of flows. While for
the 2-phase flow, the mean density is about 2 cm$^{-3}$, its value
is about 3.5 cm$^{-3}$ for the barotropic flow and  6 cm$^{-3}$
for the isothermal case. This is a natural consequence of the average 
temperature being larger in the 2-phase case than in the barotropic flow
which itself has a larger temperature than the isothermal flow.  Indeed,
since the temperature in our isothermal run is uniformly low compared
to the dynamical pressure, the
interclump gas cannot  confine significantly the clumps which are 
therefore  re-expanding once the shock that has compressed them, 
has decayed. 
For this reason, the density of the interclump gas is larger in the isothermal case 
than in the 2-phase case.

\subsection{Distribution of Mach numbers with density}

\begin{figure}
\includegraphics[width=9cm,angle=0]{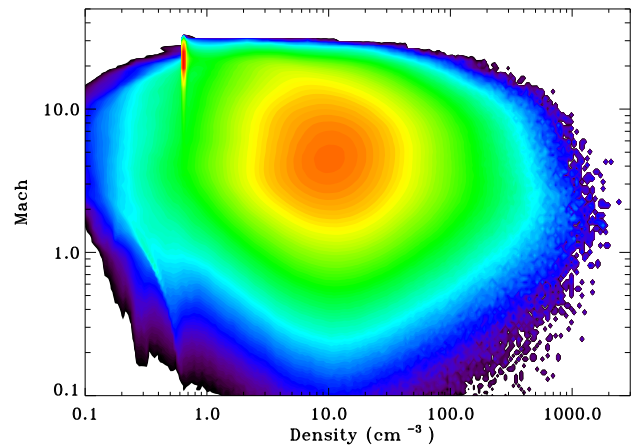}
\caption{Mass Distribution   in  the  density-Mach number  plan  for  the
isothermal  flow.  The mass excess around  $n=0.8$ cm$^{-3}$ and  ${\cal M} \simeq 15-20$
correspond to the input flow.}
\label{hist2Diso}
\end{figure}

\begin{figure}
\includegraphics[width=9cm,angle=0]{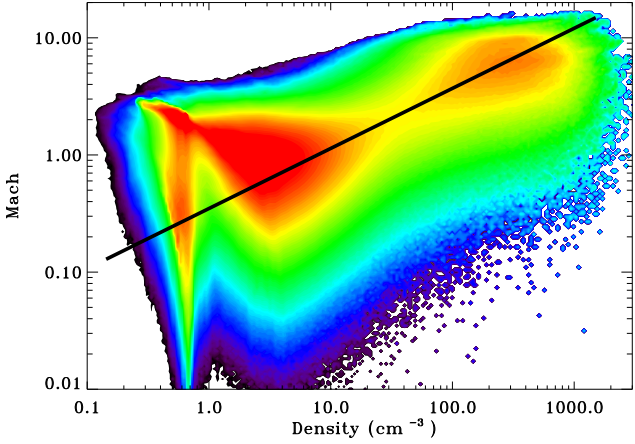}
\caption{Mass distribution in the density-Mach number plan for the bistable turbulent flow with $\epsilon=1$. The black 
line corresponds to ${\cal M} \propto \rho^{0.5}$.}
\label{hist2Dbi}
\end{figure}

Figures~\ref{hist2Diso} and \ref{hist2Dbi}  show the mass distribution
in the  density-Mach number  plan for the  2-phase and  the isothermal
flow respectively.   The isothermal case  is very similar to  what has
been previously reported  by Kritsuk et al. (2007) and Federrath et al. (2009).  
The Mach number is
essentially  not correlated  with the  density, with  a  possible weak
anti-correlation.

The two phases can clearly be seen in Fig.~\ref{hist2Dbi} and the Mach
number   varies  with  the   density  as   about  ${\cal   M}  \propto
\rho^{0.5}$. This appears to be  roughly consistent with the idea that
the  velocity  dispersion weakly  depends  on  the  density while  the
temperature is  roughly proportional  to $1/\rho$. This  last relation
naturally follows  from a nearly isobaric model.

Although  the  two   distributions  are  significantly  different,  an
interesting question is whether the cold gas itself in the 2-phase run
behave, thermally  but also  dynamically, as isothermal  or polytropic
gas. As  already discussed  in the previous  section and  also evident
from Fig.~\ref{hist2Dbi}, this  is obviously not the case  for the low
density part  of the dense  gas distribution. The question  of whether
the dense clumps,  have similar properties for both  types of flows is
investigated in the next section.

\begin{table}
\caption{Fraction of unstable and cold gas as a function of the shear} 
\label{table1}     
\centering                      
\begin{tabular}{|c| c c c| }         
\hline 
         & Cold           & Cold  & Mean value of              \\
         & + Unstable gas &       & the shear (Myrs$^{-1}$)             \\
\hline 
$\epsilon=1$     &          0.570           &    0.121  & 12.4   \\
$\epsilon=2$     &          0.569           &    0.039  & 15.0  \\
$\epsilon=4$     &          0.358           &    0.002  & 23.4  \\
\hline 
\end{tabular}
\end{table}

\begin{table}
\label{table2}     
\caption{Parameters used for the fit of the density pdf} 
\centering                      
\begin{tabular}{|c| c c c | }         
\hline 
         & $\bar{n}$ & ${\cal M}$  & b              \\
\hline 
2-phase     & 280   &  2.48  & 0.26     \\
Isothermal     & 6.5   &  3.49    & 0.33           \\
polytropic      & 3.6   &  2.95   & 0.4  \\
\hline 
\end{tabular}
\end{table}

%
%
%
%
%
%

\section{Properties of CNM structures}
In this  section, we examine  the properties of the  dense structures.
As in AH05, they are extracted  by a simple clipping algorithm using a
density threshold, $n_s$.  In the  2-phase case, 3 values of $n_s$ are
explored, namely 10, 30 and 100 cm$^{-3}$.  While the first value lies
in the thermally unstable domain,  the two others correspond to gas in
the  cold  phase.    As  for  the  previous  section,   we  also  show
corresponding  results for  isothermal  gas.  In  the isothermal  runs
however,  the  value  $n_s=10$  cm$^{-3}$  is too  low  to  allow  for
structure extraction and therefore only $n_s=30$ and 100 cm$^{-3}$ are
used. This can  be clearly seen from Fig.~\ref{pdf2}  which shows that
the density peak is close to 10 cm$^{-3}$.

Let us recall that we present the structure properties obtained in the
$\epsilon=1$  case  for  the  2-phase  run and  $\epsilon=2$  for  the
isothermal run.  The reason is that  there are more  structures in the
$\epsilon=1$ than $\epsilon=2$ case giving better statistics while the
distributions  are otherwise similar.   The corresponding  figures for
the 2-phase $\epsilon=2$ case are shown in the appendix.

\subsection{Relevance of comparison with observations}
The question of with which set of data our results should be compared
is not entirely straightforward. Strictly speaking, the 2-phase simulations should 
be compared with HI data such as the one obtained in the Millenium
survey (Heiles \& Troland 2003). This comparison which was the main 
purpose of the study performed by Hennebelle et al. (2007), is 
however restricted. The reason is that the distance of HI clouds 
is usually unknown because HI is ubiquitous within the Galaxy. 
Thus, in particular, size and mass spectrum cannot be easily computed.
Extragalactic studies such as the one performed by Kim et al. (2007)
can get rid of this difficulty, however, the scales that one can probe 
in nearby galaxies is much larger than the ones tackled in our 
simulations.

On the other hand, our results are not restricted to 2-phase flows since 
isothermal and barotropic cases are also considered. These thermal approximations 
are thought to be fairly reasonable to describe the dense part of 
molecular clouds (i.e. denser than 10$^3$ cm$^{-3}$) and have been used 
in many studies. Thus it is both interesting and justify to compare 
the statistics inferred from these numerical simulations with the 
statistics which have been observationally inferred for molecular clouds.

Finally, the nature of the interclump medium within molecular clouds
is still very uncertain. Williams et al. (1995) studied in detail
the Rosette molecular cloud and conclude that in this cloud, the 
interclump medium is a mixture of atomic gas 
(the density, 2-4 cm$^{-3}$, they quote, traditionally 
corresponds to either warm or thermally unstable HI gas as seen in Fig. 4)
and very diffusive H$_2$. Recent theoretical studies (e.g. Hennebelle et al. 
2008, Heitsch et al. 2008, Banerjee et al. 2009) attempting to form 
molecular gas from the diffusive atomic gas, argue that WNM persists between
the clumps inside the molecular clouds.  If this assumption is correct, 
this would imply that molecular clouds are also 2-phase 
objects, with dense molecular gas embedded into a warm and diffuse atomic phase.
For this reason, we believe that with due cautions, comparisons between our 
2-phase results and some statistics inferred for molecular clouds are also 
relevant.

\subsection{Clump mass spectrum}

\begin{figure}
\includegraphics[width=9cm,angle=0]{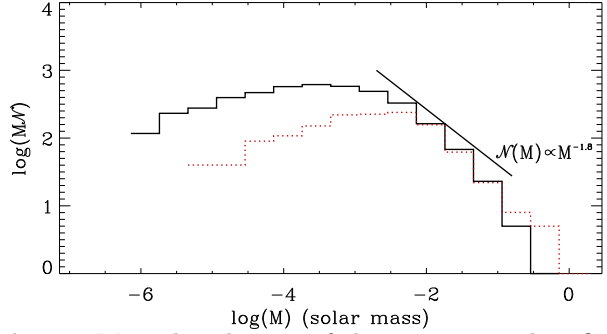}
\caption{Mass distribution of the structures identified in the  2-phase simulations (A1) with 
a density threshold $n_c=10$ cm$^{-3}$. Solid black line shows the 1200$^3$ simulation
while red dotted line shows the 600$^3$ simulation. The black straight line shows 
a clump mass spectrum $dN/dM \propto M^{-1.8}$.}
\label{mass_spect_2p_10}
\end{figure}

\begin{figure}
\includegraphics[width=9cm,angle=0]{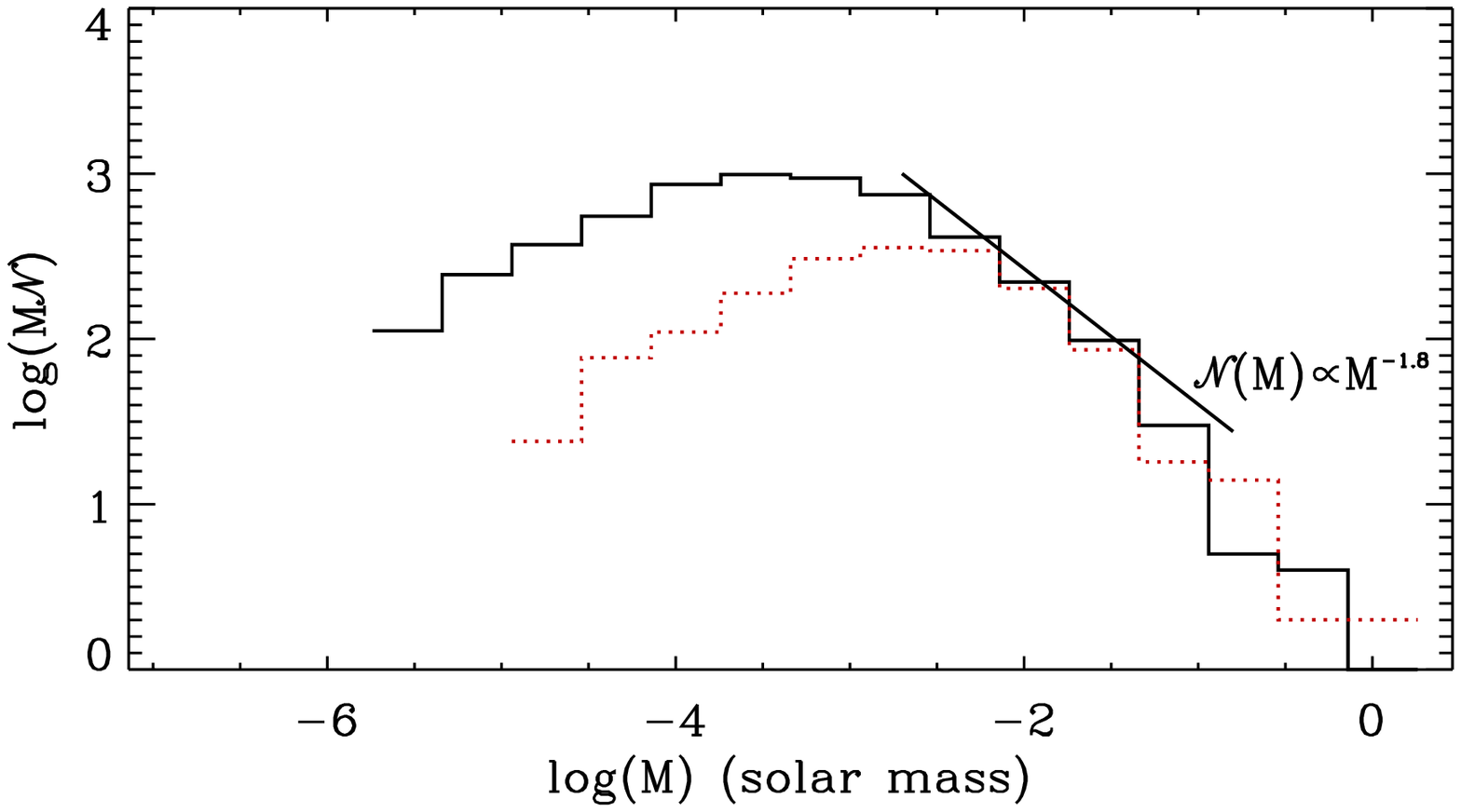}
\caption{Same as Fig.~\ref{mass_spect_2p_10} for 
a density threshold $n_c=30$ cm$^{-3}$.}
\label{mass_spect_2p_30}
\end{figure}

\begin{figure}
\includegraphics[width=9cm,angle=0]{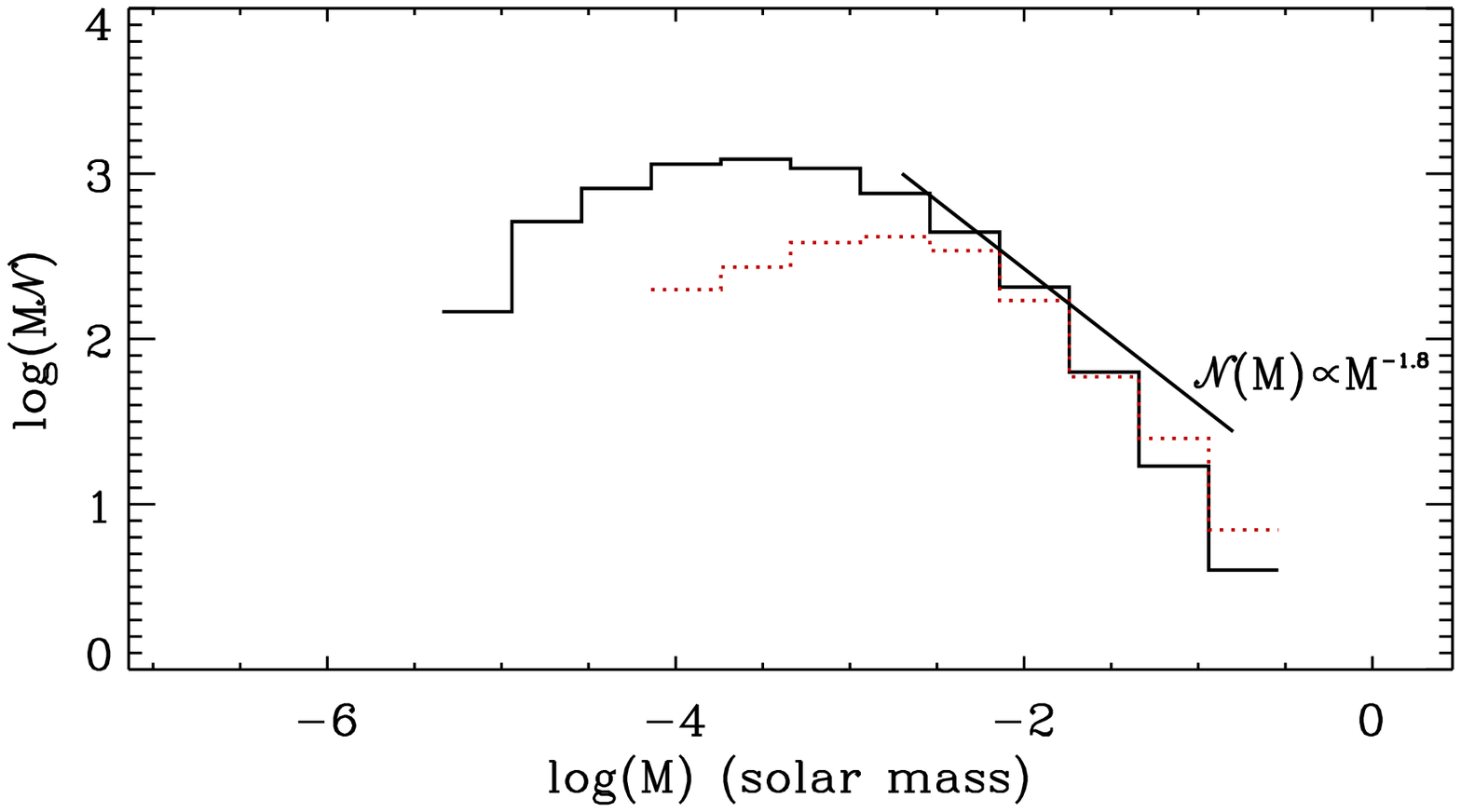}
\caption{Same as Fig.~\ref{mass_spect_2p_10} for 
a density threshold $n_c=100$ cm$^{-3}$.}
\label{mass_spect_2p_100}
\end{figure}

\begin{figure}
\includegraphics[width=9cm,angle=0]{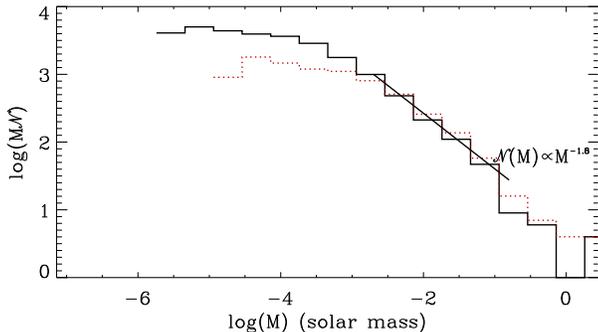}
\caption{Same as Fig.~\ref{mass_spect_2p_10} for the isothermal simulation and
a density threshold $n_c=30$ cm$^{-3}$.}
\label{mass_spect_iso_30}
\end{figure}

Figures~\ref{mass_spect_2p_10}-\ref{mass_spect_2p_100}  show  the mass
spectrum for  the structures extracted  from the $600^3$  and $1200^3$
2-phase simulations  and with  a density threshold  $n_s$ respectively
equal to  10, 30 and  100 cm$^{-3}$.  While  the low mass part  of the
distributions ($<10^{-3}$  solar masses)  is spoiled by  the numerical
resolution, the high mass part  ($>0.1$ solar masses) suffer from poor
statistics due to the finite size  of the box. However, as can be seen
by comparing the results of  the $600^3$ and the $1200^3$ simulations,
one can see  that numerical resolution seems to  be nearly reached for
intermediate  masses. For  the   three  density  thresholds,  the  mass
spectrum in this range  of masses, approximately follows ${{\cal N}(M)
\propto  M^{-1.8}}$. It  is therefore  very similar  to what  has been
inferred in the 2D simulations  presented in HA07.  It is also similar
to the mass spectrum inferred by Dib et al. (2008).

Figure~\ref{mass_spect_iso_30}   shows  the   mass  spectrum   in  the
isothermal  case for  $n_c=30$ cm$^{-3}$.  The distribution  is rather
similar to the 2-phase case except interestingly that more small scale
structures are found. We interpret  this as being a consequence of the
effective (or numerical) Field  length preventing the formation of CNM
structures of  size too close to  the mesh.  As for  the 2-phase case,
the mass spectrum  between $10^{-3}$ and $10^{-1}$ solar  masses, is a
powerlaw whose exponent is close to $1.8$.

\begin{figure}
\includegraphics[width=9cm,angle=0]{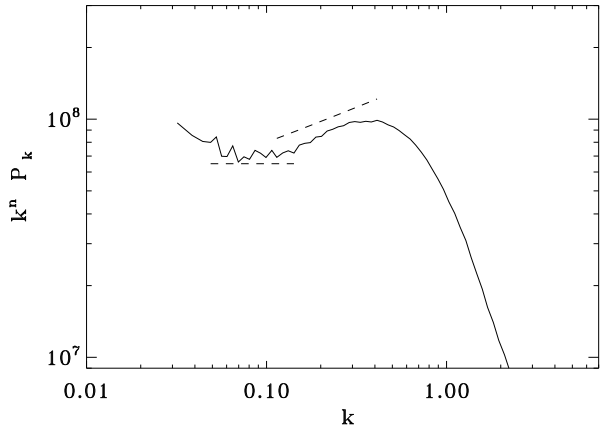}
\caption{Compensated power spectrum $P_k k^{n'}$ of the logarithm of the density with $n'=3.3$ for the weakly turbulent simulation. The two dashed horizontal  lines correspond to slopes of -3.3 and -3.1.}
\label{ps_log_2p}
\end{figure}

\begin{figure}
\includegraphics[width=9cm,angle=0]{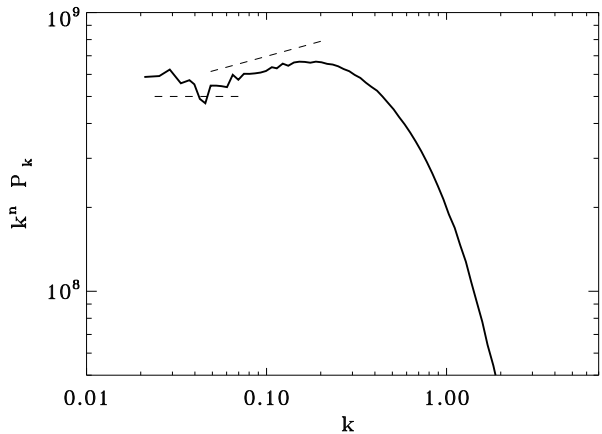}
\caption{Compensated power spectrum $P_k k^{n'}$ of the logarithm of the density with $n'=3.3$ for the isothermal simulation. The two dashed horizontal 
lines correspond to slopes of -3.3 and -3.1.}
\label{ps_log_iso}
\end{figure}

Let us recalled that in HA07, we propose a theory to 
explain the mass spectrum of the CNM structures which predicts $\gamma=16/9$ in 3D
and is therefore compatible with the numerical results. This theory,  which is based on 
the Press \& Schecter (1974) formalism and assume that the structure mass spectrum reflects the
density fluctuations arising in the trans-sonic WNM, 
has been recently extended by Hennebelle \& Chabrier (2008) to the supersonic case.
In particular,  the exponent of the structure mass spectrum is related
to the exponent of the powerspectrum of $\log(\rho)$, $n'$, through the relation:
\begin{eqnarray}
{ d N \over d M} \propto M^{-\beta}=M^{-2+{n'-3 \over 3}}.
\label{equ_mass_spec}
\end{eqnarray}
In order to  verify this relation, we have  computed the powerspectrum
of   $\log    (\rho)$.    The    2-phase   case   is    presented   in
Fig.~\ref{ps_log_2p}   while   the   isothermal   one  is   shown   in
Fig.~\ref{ps_log_iso}.  The value of  the exponent  is measured  to be
about 3.3 while as in Kritsuk et al. (2007), we identify a bottle neck
in  which  the exponent  is  slightly  shallower.  This value  of  the
exponent  is similar  to what  is reported  in Schmidt  et  al. (2009)
thought slightly smaller  since they infer a value  closer to 3.8.  We
speculate that  this maybe due to  our boundary conditions  since in a
significant fraction of  the computational box, the flow  is not fully
turbulent.

Taking $n'  \simeq 3.3$, we get  with Eq.(\ref{equ_mass_spec}), $\beta
\simeq  1.9$ which  appears to  be compatible  with the  mass spectrum
inferred from the simulations.  Taking into account the value inferred
by Schmidt et  al. (2009), i.e. $n'=3.8$, we  get $\beta \simeq 1.75$.
We also  note that the theory  presented in HA07 and  in Hennebelle \&
Chabrier (2008)  predicts that the  mass spectrum exponent  should not
depend on the density threshold  which seems to be compatible with the
results                          displayed                          in
Figs.~\ref{mass_spect_2p_10}-\ref{mass_spect_2p_100}.  Note  that with
larger values of  $n_c$, the mass spectrum we  get is not sufficiently
accurate to  obtain reliable estimate  of the exponent.   Finally, the
mass  spectra  obtained  for   different  values  of  $\epsilon$,  the
fluctuation amplitude  imposed at the boundaries, are  very similar to
the mass spectra presented above, as shown in the appendix.

It  is  remarkable that  the  mass  spectra  obtained under  different
conditions of forcing and  for different thermodynamical behaviors are
so  similar particularly  because  significant density  PDF have  been
inferred.   Indeed,  this  strongly  suggests  that  turbulence  whose
behaviour tends to be universal, is mainly responsible of the shape of
the mass spectrum as indicated by Eq.~(\ref{equ_mass_spec}).

The value of $\beta$ inferred from the simulations are also compatible
with  the value quoted  in various  observational studies  (e.g. Blitz
1993,  Heithausen  et al.  1998)  for  the  CO clumps.   Interestingly
enough, the structure masses in the sample of Heithausen et al. (1998)
are as  small as  one Jupiter mass  and therefore compatible  with the
mass   of  the   structures  formed   in  our   simulations.   Although
straightforward  comparison and therefore  conclusion should  be taken
with care at this stage, since the present work focuses on HI and does
not treat neither  the formation of H$_2$ nor the  formation of the CO
molecule, the  universality of the  mass spectrum that we  observed in
our simulations suggests that these processes likely do not affect the
mass spectrum significantly.  We also note that in a recent study, Kim
et al. (2007)  inferred similar values for the HI  clouds in the small
magellanic cloud.  The mass of these clouds  (10$^{4-6}$ solar masses)
is however much larger than the mass of our simulated clouds.

\subsection{Clump mass size relation}
In this section,  we investigate the mass-size relation.   As in HA07,
the size  is defined  by computing the  inertia matrix and  taking its
largest eigenvalue.   Note that we  have tried other choices,  like for
example  the geometric  means of the 3 eigenvalues  and  it does  not  change the  results
significantly.

\begin{figure}
\includegraphics[width=9cm,angle=0]{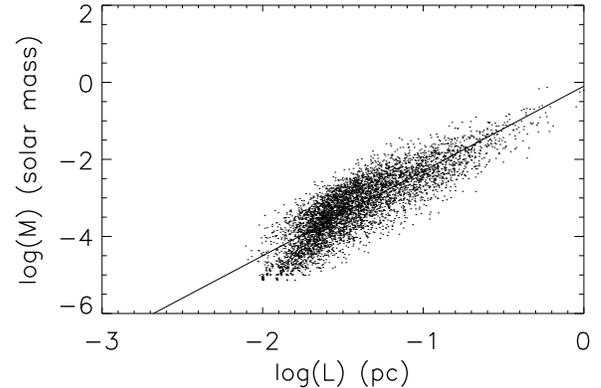}
\caption{Mass versus size relation for the CNM structures extracted from the 2-phase $1200^3$ cells simulations.}
\label{struct_2phase}
\end{figure}

\begin{figure}
\includegraphics[width=9cm,angle=0]{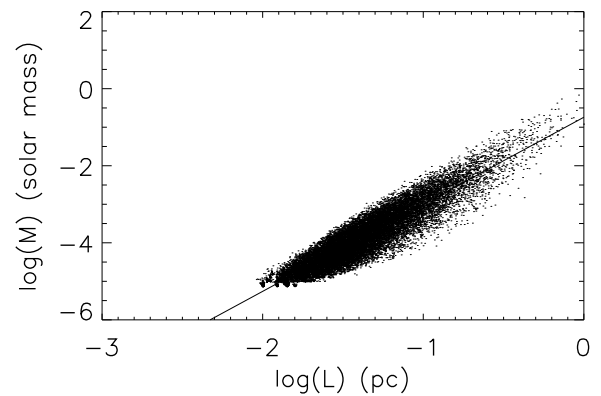}
\caption{Mass versus size relation for the structures extracted from the isothermal $1200^3$ cells  simulations.}
\label{struct_iso}
\end{figure}

Figure~\ref{struct_2phase} shows the mass versus size relation for the CNM structures extracted from the 
$1200^3$ cells  simulations for a value of $n_c$=30 cm$^{-3}$. Structures obtained with 
$n_c=$ 10 and 100 cm$^{-3}$ show very similar behaviors and are therefore
not displayed here for conciseness. This is also the case for the structures 
formed in simulations with a larger fluctuation amplitude, $\epsilon=2$ as can be seen in the 
appendix.
The straight line represents a linear fit of the whole
distribution. The slope is about 2.2. As can be seen, it represents well the distribution
over the whole range of masses.

Figure~\ref{struct_iso} shows  the mass  versus size relation  for the
structures of the isothermal simulation.  The slope is about 2.25 and
therefore  very  similar  to  the  2-phase case.   The  shape  of  the
distribution is slightly different from  the 2-phase case and tends to
be  closer to  a straight  line.  The  similar behaviour  obtained for
isothermal and  2-phase flows, again  suggests that turbulence  is the
main  mechanism responsible of  the structure's  shape.  The  slope of
2.25 derived from our simulation, is very close to what is reported in
Kritsuk et  al. (2007) and Federrath  et al. (2009)  although they use
different  definitions and algorithms than ours which  is  more
appropriate to compare with observations.

Finally, it  is worthwhile to  recall that the  value of 2.3  has been
obtained by Heithausen et al. (1998)  and Kramer et al. (1998) for the
CO  molecular clumps.  Using a  larger  sample of  data, Falgarone  et
al. (2004) also  conclude that $M \propto L^{2.3}$  is more compatible
with the data  than the relation $M \propto  L^{2}$, originally quoted
by Larson (1981).

\subsection{Velocity dispersion within structures}

\begin{figure}
\includegraphics[width=9cm,angle=0]{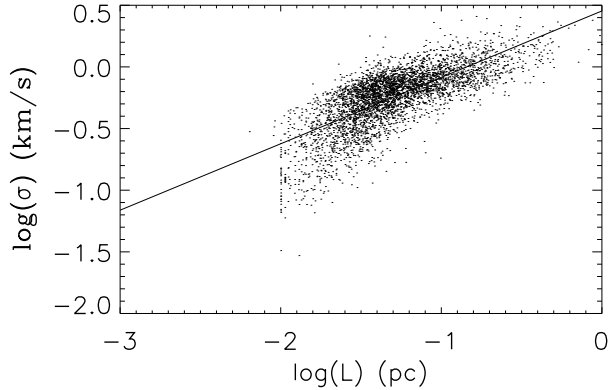}
\caption{Velocity dispersion versus size relation of the CNM structures extracted from the 2-phase $1200^3$ cells simulations for $n_c=10$ cm$^{-3}$.}
\label{vel_2p_10}
\end{figure}

\begin{figure}
\includegraphics[width=9cm,angle=0]{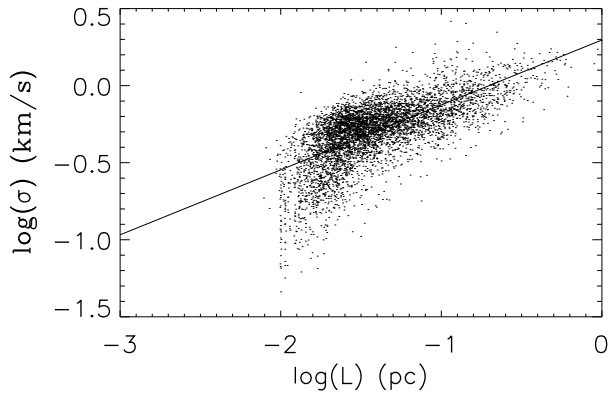}
\caption{Same as Fig.~\ref{vel_2p_10} for $n_c=30$ cm$^{-3}$.}
\label{vel_2p_30}
\end{figure}

\begin{figure}
\includegraphics[width=9cm,angle=0]{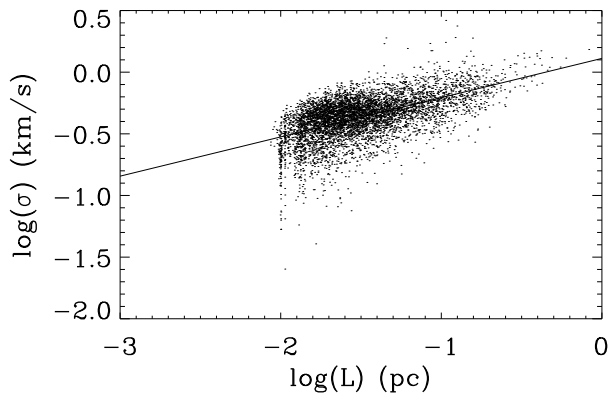}
\caption{Same as Fig.~\ref{vel_2p_10} for $n_c=100$ cm$^{-3}$.}
\label{vel_2p_100}
\end{figure}

\begin{figure}
\includegraphics[width=9cm,angle=0]{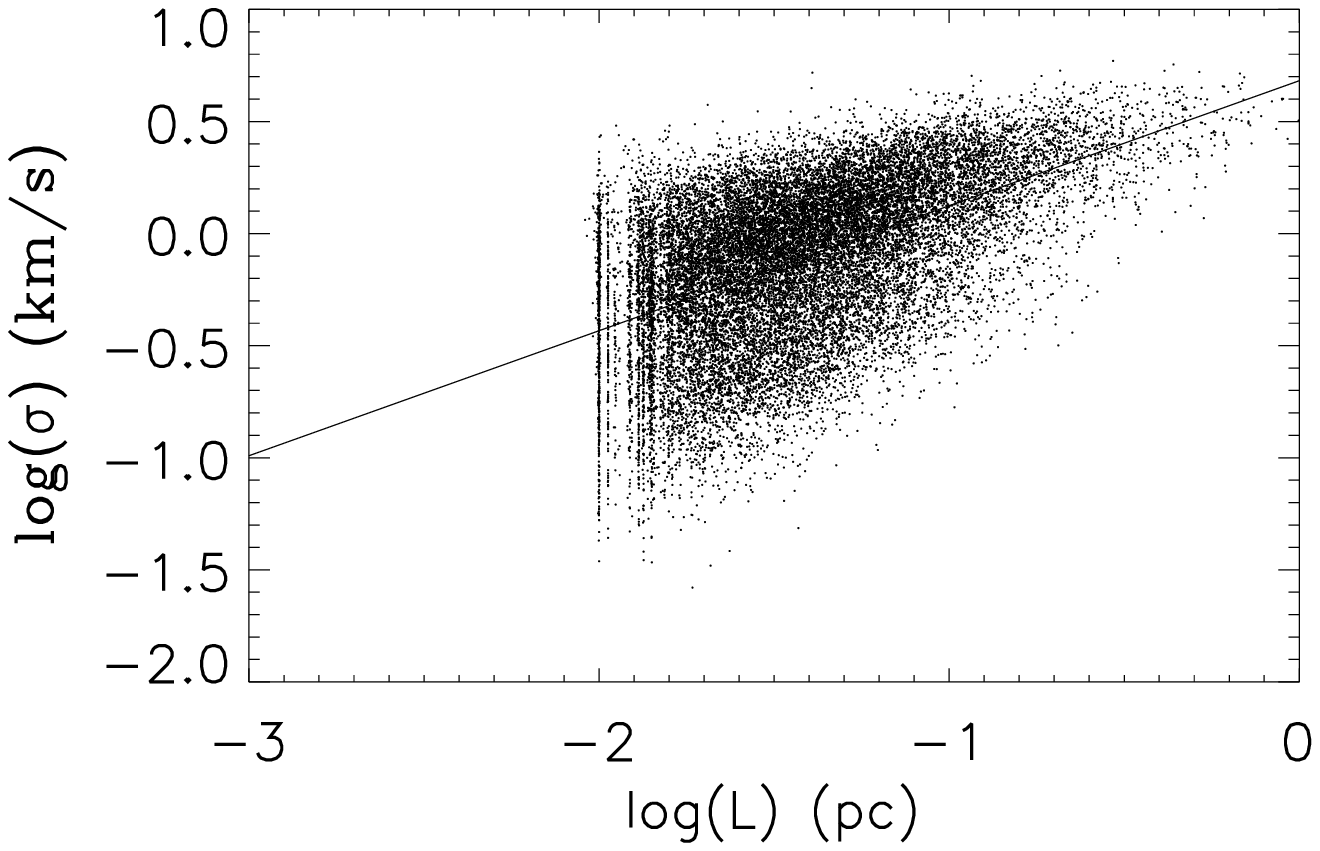}
\caption{Velocity dispersion versus size relation of the structures extracted from the $1200^3$ cells isothermal  simulations  for $n_c=30$ cm$^{-3}$.}
\label{vel_iso_30}
\end{figure}

\begin{figure}
\includegraphics[width=9cm,angle=0]{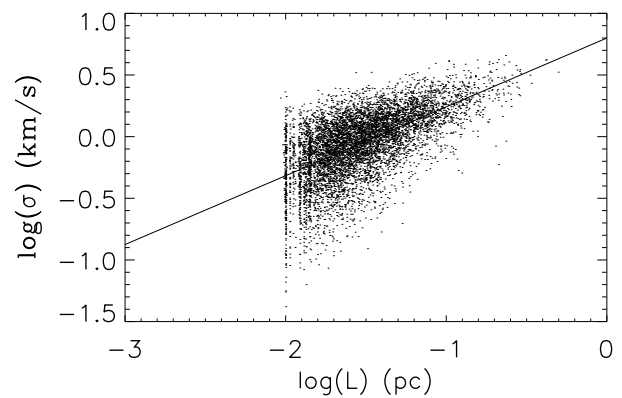}
\caption{Velocity dispersion versus size relation of the structures 
extracted from the $1200^3$ cells isothermal  simulations  for $n_c=100$ cm$^{-3}$.}
\label{vel_iso_100}
\end{figure}

Figures~\ref{vel_2p_10}-\ref{vel_2p_100}   show  the   total  internal velocity
dispersion  versus size  for the  CNM structures,  i.e.  $\sqrt{\delta
V_x^2 +  \delta V_y ^ 2  + \delta V_z ^  2}$ as a function  of $L$ for
$n_c$ equal  to respectively 10, 30  and 100 cm$^{-3}$  in the 2-phase
case, where $\delta V$ is the velocity with respect to the structure
mean velocity.  The straight line represents  a linear fit of the distribution.
While the relation $\sigma \simeq 3.3$ km s$^{-1}$ ($L$/1 pc)$^{0.53}$
is  inferred for  $n_c$=10 cm$^{-3}$,  we find  $\sigma \propto  2$ km
s$^{-1}$  ($L$/1  pc)$^{0.43}$  for  $n_c$=30  cm$^{-3}$  and  $\sigma
\propto 1.2$ km s$^{-1}$ ($L$/1 pc)$^{0.33}$ for $n_c$=100 cm$^{-3}$.
It is interesting to note that for the three values of $n_c$, 
$n_c \sigma^2 (L)$ is roughly the same. 
That is 10$\times$3.3$^2 \simeq 30 \times 2^2 \simeq 100 \times 1.2^2$.
This suggests that kinetic 
energy tends to be equally distributed in the volume rather than in the mass
of the flow. Moreover, it is also interesting to note that this dynamical or 
turbulent pressure is of the same orders as the thermal pressure and incoming ram 
pressure.  Indeed, for  $n_c$=100 cm$^{-3}$ the thermal pressure 
of the clump is about $10^4 k_b $ K cm$^{-3}$ since the temperature is 
about 50K-100K while the turbulent pressure is about  
$ 2 \times 10^4 k_b (L/1 pc)^2 $ K cm$^{-3}$, 
as calculated previously  the ram pressure of the incoming flow 
is about $ 2 \times 10^4 k_b$ K cm$^{-3}$. 
Note that in the velocity dispersion, all the motions have been counted 
irrespectively of the fact that their could be solenoidal, 
converging or diverging modes. Straighforward interpretation
of this velocity dispersion as a turbulent pressure is therefore 
not entirely obvious and certainly not accurate within more than 
a factor of a few.

Figures~\ref{vel_iso_30}-\ref{vel_iso_100}  show  the  total  velocity
dispersion  versus size  for $n_c$  equal to  respectively 30  and 100
cm$^{-3}$ in  the isothermal case.  In  both cases, we  find a similar
relation of about $\sigma \simeq 5$ km s$^{-1}$ ($L$/1 pc)$^{0.55}$
although for $n_c=100$ cm$^{-3}$, there are 
no clump of size bigger than 0.5 pc and very few with size bigger than 
0.3 pc. 
Interestingly, this is a different behaviour than for the 2-phase case
since the energy is now more concentrated in dense regions.

These results suggest that 2-phase and isothermal flows behave somehow
differently.   In particular, since  the velocity  dispersion is  about 2  to  3 times
higher  while the  forcing is  identical  in both  cases, 
 energy is  more efficiently injected into the dense clumps in isothermal flows
than it is in 2-phase  flows.  Three,  not
exclusive,  interpretations  sound  possible  to us.   First,  in  the
2-phase  flow,  energy is  efficiently  radiated  away, in  particular
during the transition  between the warm and the  cold phase, where the
effective  adiabatic exponent  is  negative.  Second,  in the  2-phase
flows,  the clumps  are  long living  since  they are  bounded by  the
external pressure. Thus,  after a crossing time most  of the turbulent
motions, initially  present in the structure have  dissipated. This is
different  from the  isothermal case  in which  the structures  do not
survive   more   than  an   expansion   time   since   they  are   not
confined by the external thermal pressure. 
Third, since the dense structures possess sharp edges in the
2-phase flow while in the isothermal one, the transition between dense
and  rarefied  gas  is  continuous,  the coupling  between  the  dense
material and the external medium is  weaker for the former than for the
latter. In  particular, it  sounds likely that  sound waves  should be
more efficiently reflected by the stiff discontinuities in the 2-phase
case. Making a quantitative estimate  of these effects is difficult at
this stage.

The value of the exponent 0.3-0.5 in this relation is again remarkably
similar  to what  is  inferred observationally  in  our galaxy  (e.g.,
Larson  1981). It  is also  in good  agreement with  the index  of the
velocity powerlaw,  $P(v)$, being in the  range 11/3 to  4 as inferred
from many numerical  simulations (e.g., Kritsuk et al.  2007).  Let us
recall that if $\sigma \propto L^\eta$ and $P(v) \propto k^{-\alpha}$,
then $\alpha -3 = 2 \eta$.
 
We note however, that while the slope of the velocity-size relation is
approximately   correct,  its  value   itself  may be  too
high. Indeed one infers from  observations of the CO clumps, the value
1 km  s$^{-1}$ for the  line width which  corresponds to about  0.4 km
s$^{-1}$  for the  velocity dispersion  along  the line  of sight  and
therefore  to about  $\simeq$0.8 km  s$^{-1}$ for  the  total velocity
dispersion assuming isotropy which is  smaller than what we infer here
by  a factor  1.5  to 3  in  the 2-phase  case and  about  5-6 in  the
isothermal one. Note that since the density of the CO clumps is of the
order of 3000 cm$^{-3}$, the comparison cannot be more than indicative
at this  stage. We  also note that  the velocity  dispersion decreases
with increasing density in our 2-phase simulations suggesting that the
velocity dispersion could be very similar to what has been inferred by
Larson  (1981)  for denser  clumps.   Since  the velocity  dispersion,
directly  depends on  the  forcing, another  possibility  is that  the
forcing is a little too strong and should be reduced to better fit the
mean ISM conditions.

\section{Conclusion}
In this paper, we perform  3D high resolution numerical simulations of
converging flows  using either a standard  interstellar atomic cooling
function,  an isothermal  or a  polytropic equation  of state  with an
adiabatic index of $\gamma=0.7$.

We  investigate  the density  PDF  and  Mach number-density  relation,
showing  that, as expected,  the thermal  behaviour of  the gas  has a
drastic  influence.   While as  previous  authors,  we  find that  the
isothermal runs tend to  produce lognormal density distributions, when
the adiabatic  exponent is smaller  than one, namely  $\gamma=0.7$, we
find  that  the  density  distribution  follow  a  powerlaw  for  high
densities  confirming  the  result  of  Passot  \&  V\'azquez-Semadeni
(1998).  The 2-phase  case is very different thought  the high density
part of the distribution could  reasonably be described by a lognormal
distribution.   We stress  that higher  numerical resolution  may well
change this conclusion.  However,  the distribution of the low density
(say  in the range  10-300 cm$^{-3}$)  cold gas  departs significantly
from it, making unclear the  use of the lognormal density distribution
an adequate choice for molecular clouds.

We  compute  the mass  spectrum  of  the  clumps for  various  density
thresholds,  as  well  as  the  mass-size relation  and  the  internal
velocity dispersion-size  relation.  We find that  the 3 distributions
agree  with  the  statistics  inferred  for  the  interstellar  clouds
(e.g.  Heithausen et  al. 1998)  reasonably well.
In particular, the 3 distributions are well fitted by a powerlaw whose
exponent values are close to the observed ones.
   While the  first 2
appear to be  reasonably similar for isothermal and  2-phase flows, we
find  that the  internal velocity  dispersion is  about 2  to  3 times
larger for the  clumps of the isothermal simulation  than for the ones
of  the   2-phase  case,  suggesting  that  energy is less 
efficiently injected into the  dense clumps in 
2-phase   flows  than it is in isothermal ones.

Altogether, these results confirm the  claim made in HA07 that 2-phase
flows behave differently than isothermal flows. Although it appears to
us that  higher resolution simulations should be  performed to further
assess this  conclusion, the present  study suggests that  the 2-phase
nature  of the  flow may  have significant  implications even  for the
physics of high density gas.

\section{Acknowledgments}
We thank Christoph Federrath for a critical reading of the manuscript
as well as an anonymous referee for his/her help in clarifying the original 
version of this work significantly. This work was granted access to the HPC resources
of CCRT and CINES under the allocation x2009042204 and x2009042036
made by GENCI (Grand Equipement National de Calcul Intensif).
EA acknowledge support from the French ANR through the SiNERGHy 
project, ANR-06-CIS6-009-01.

\appendix

\section{Statistics for clumps in the $\epsilon=2$ simulation}
Here, for completeness, we give the various statistics of the clumps formed in the simulation 
performed with $\epsilon=2$.

\begin{figure}
\includegraphics[width=9cm,angle=0]{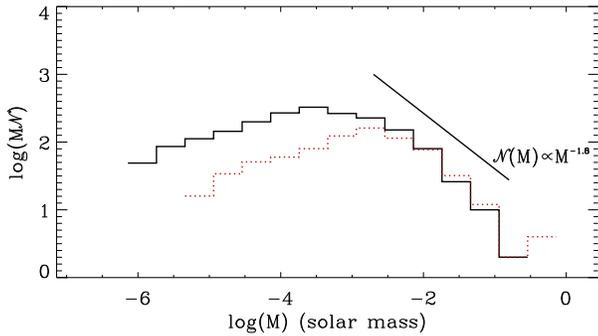}
\caption{Mass distribution of the structures identified in the  2-phase simulations (A2) with 
a density threshold $n_c=10$ cm$^{-3}$. Solid black line shows the 1200$^3$ simulation
while red dotted line shows the 600$^3$ simulation. The black straight line shows 
a clump mass spectrum $dN/dM \propto M^{-1.8}$.}
\label{mass_spect_2p_10_A2}
\end{figure}

\begin{figure}
\includegraphics[width=9cm,angle=0]{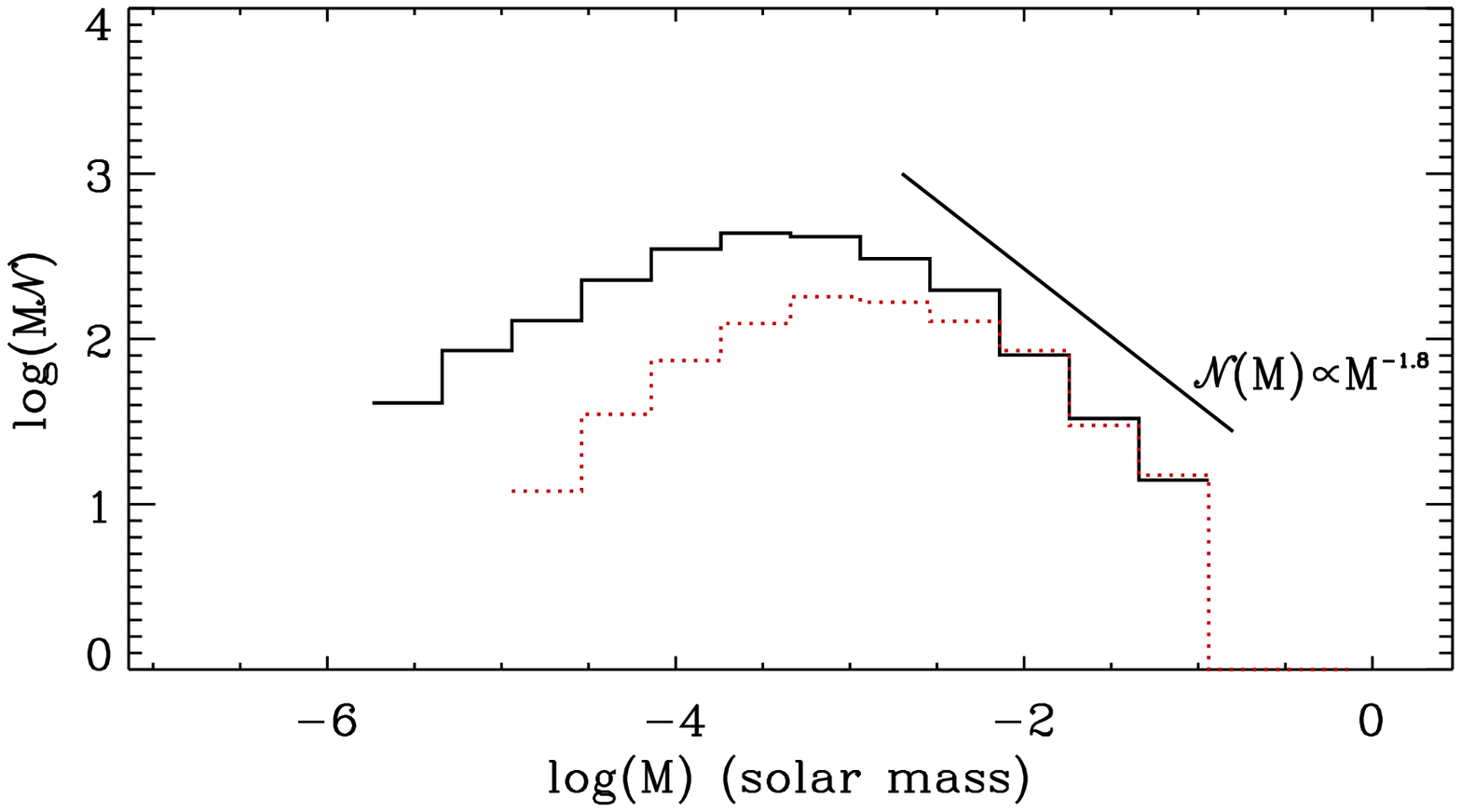}
\caption{Same as Fig.~\ref{mass_spect_2p_10_A2} for 
a density threshold $n_c=30$ cm$^{-3}$.}
\label{mass_spect_2p_30_A2}
\end{figure}

\begin{figure}
\includegraphics[width=9cm,angle=0]{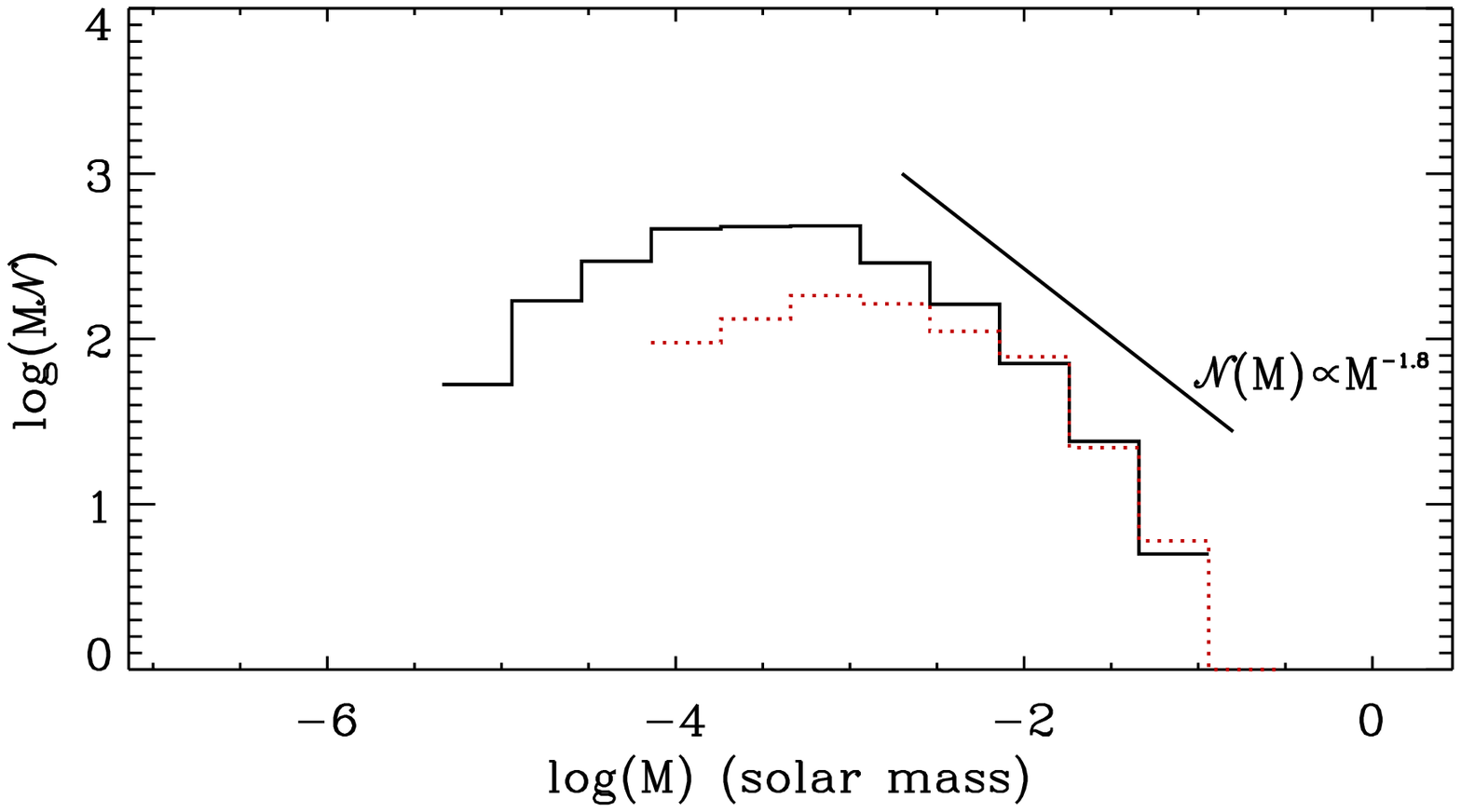}
\caption{Same as Fig.~\ref{mass_spect_2p_10_A2} for 
a density threshold $n_c=100$ cm$^{-3}$.}
\label{mass_spect_2p_100_A2}
\end{figure}

\begin{figure}
\includegraphics[width=9cm,angle=0]{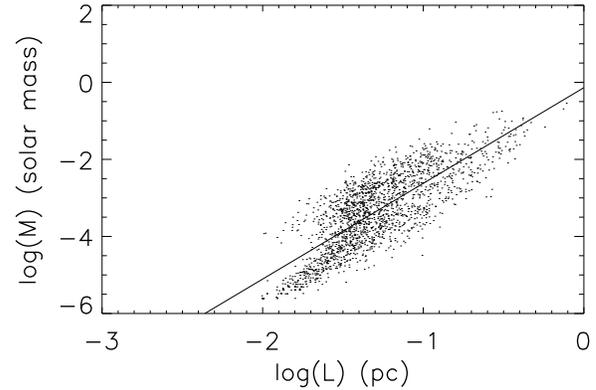}
\caption{Mass versus size relation for the CNM structures extracted from the $1200^3$ cells simulations with $\epsilon=2$
 for $n_c=10$.}
\label{mass_L_10_A2}
\end{figure}

\begin{figure}
\includegraphics[width=9cm,angle=0]{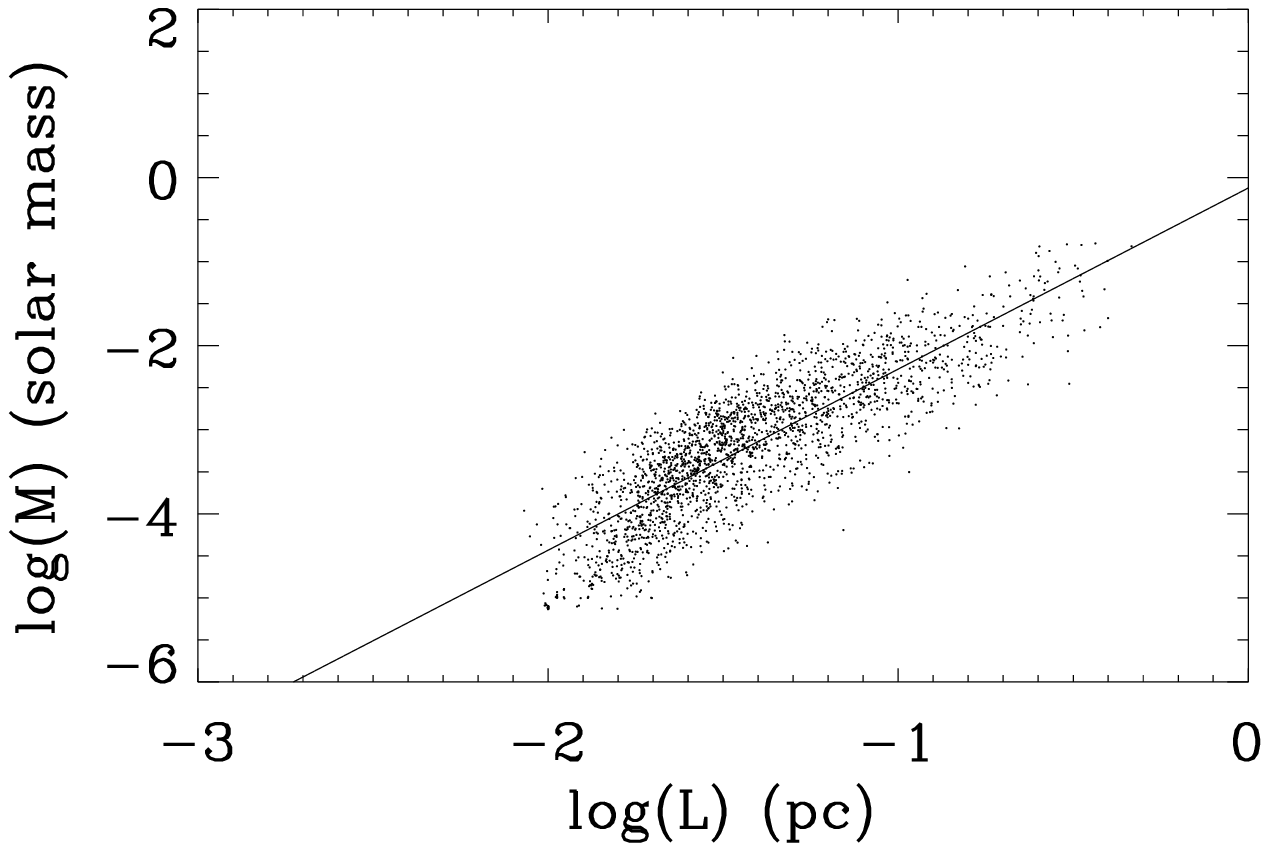}
\caption{Same as Fig.~\ref{mass_L_10_A2} for $n_c=30$ cm$^{-3}$.}
\label{mass_L_30_A2}
\end{figure}

\begin{figure}
\includegraphics[width=9cm,angle=0]{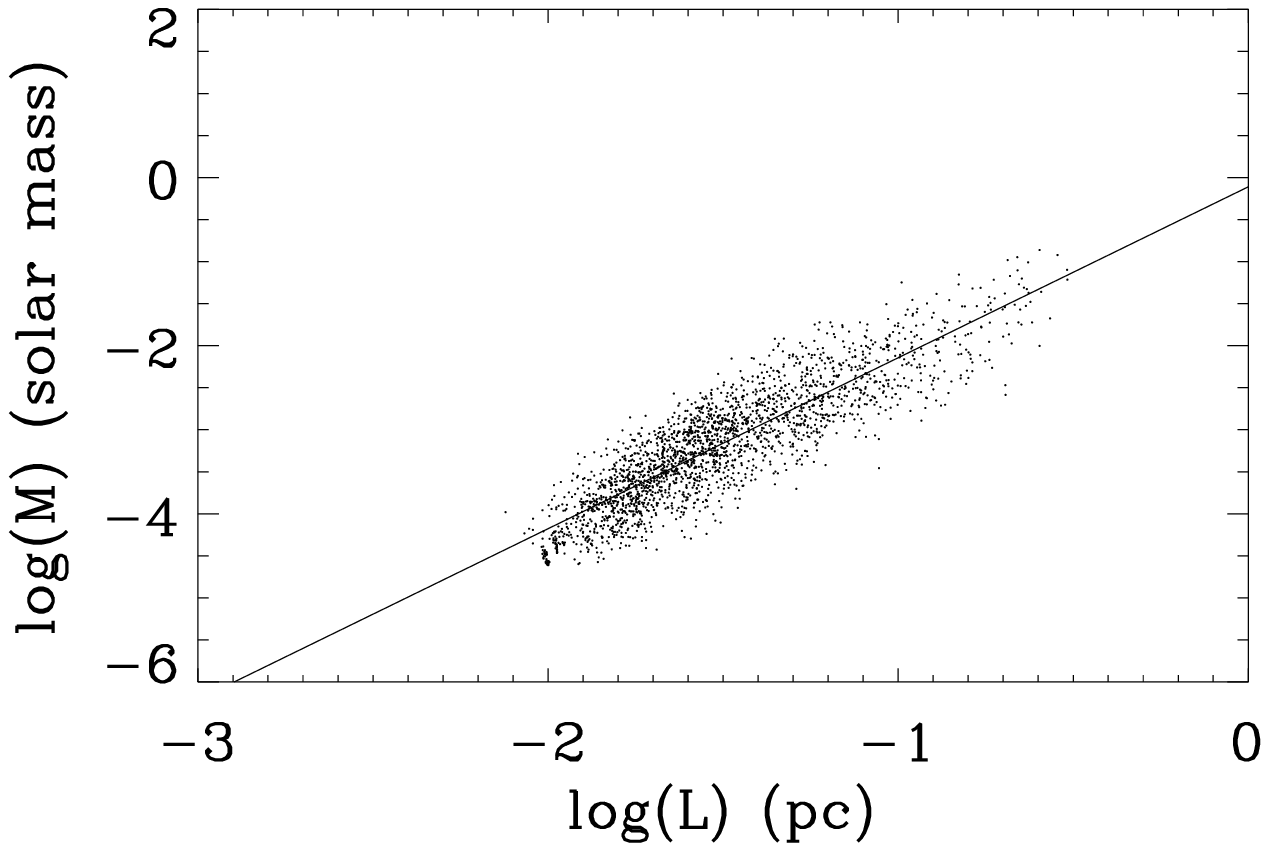}
\caption{Same as Fig.~\ref{mass_L_10_A2} for $n_c=100$ cm$^{-3}$.}
\label{mass_L_100_A2}
\end{figure}

\begin{figure}
\includegraphics[width=9cm,angle=0]{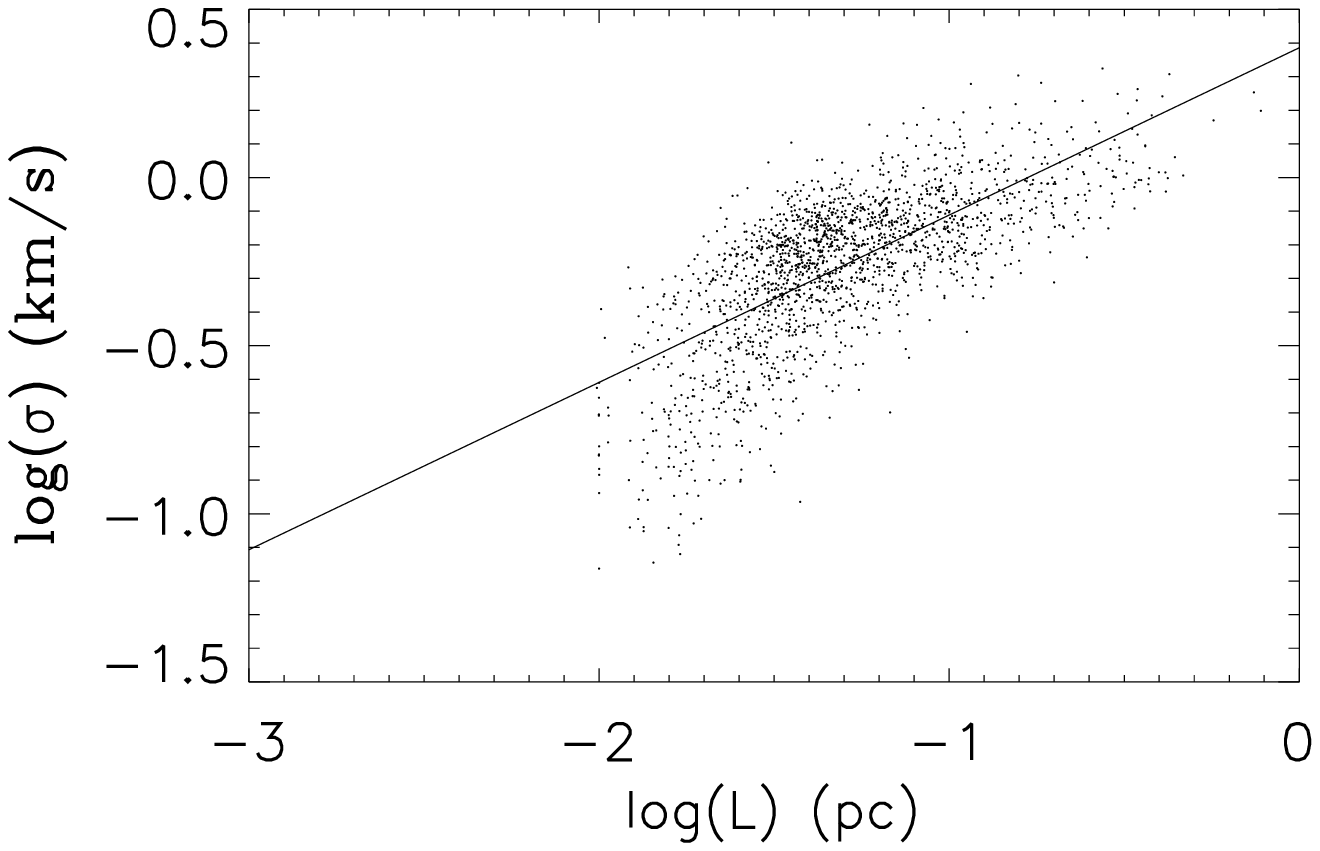}
\caption{Velocity dispersion versus size relation of the CNM structures extracted from the 2-phase $1200^3$ cells simulations for $n_c=10$ cm$^{-3}$.}
\label{vel_2p_10_A2}
\end{figure}

\begin{figure}
\includegraphics[width=9cm,angle=0]{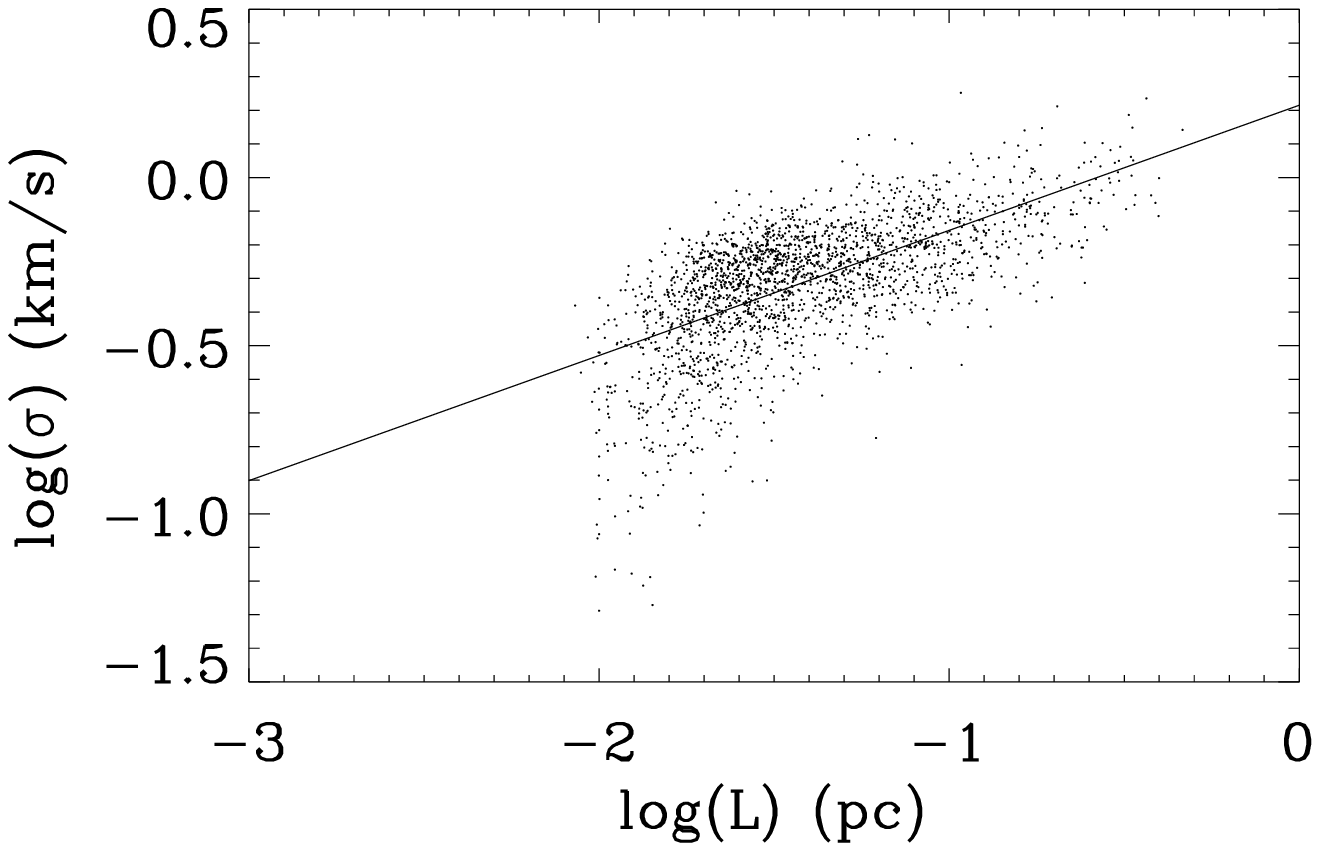}
\caption{Same as Fig.~\ref{vel_2p_10_A2} for $n_c=30$ cm$^{-3}$.}
\label{vel_2p_30_A2}
\end{figure}

\begin{figure}
\includegraphics[width=9cm,angle=0]{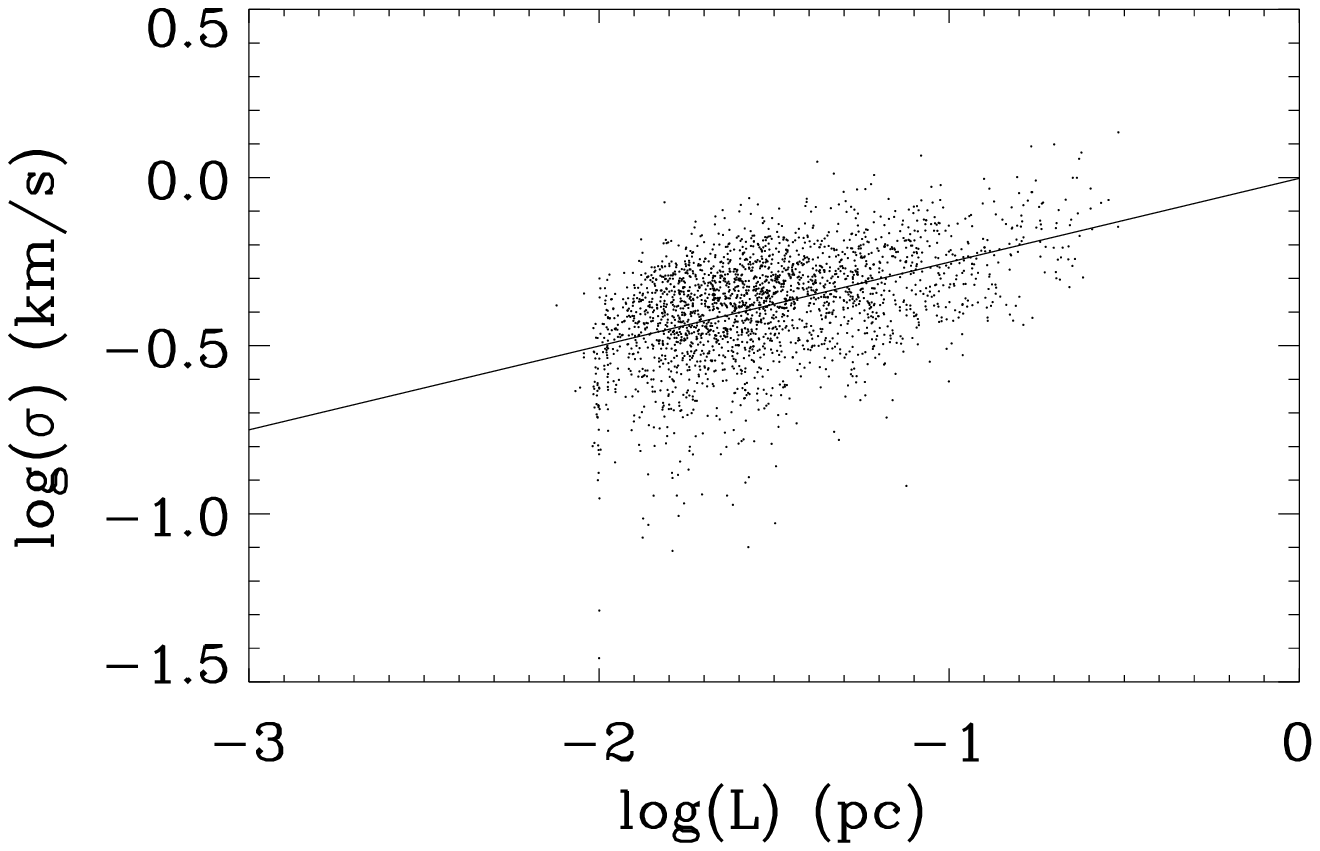}
\caption{Same as Fig.~\ref{vel_2p_10_A2} for $n_c=100$ cm$^{-3}$.}
\label{vel_2p_100_A2}
\end{figure}

\end{document}